\documentclass[]{aa} 
\usepackage{graphicx} 
\usepackage{txfonts} 
\usepackage{color}

\begin{document} 

\title{Gamma-ray activity of Seyfert galaxies and constraints on hot accretion flows}   

\author{Rafa{\l} Wojaczy\'nski \inst{1}, Andrzej Nied\'zwiecki \inst{1},
Fu-Guo Xie\inst{2}, Micha{\l} Szanecki \inst{1}}  
\institute{Department of Astrophysics, University of \L \'od\'z, Pomorska 149/153, 
90-236 \L \'od\'z, Poland\\  
\email{rafal.wojaczynski@wp.pl, niedzwiecki@uni.lodz.pl, mitsza@uni.lodz.pl}
\and
Key Laboratory for Research in Galaxies and Cosmology, Shanghai
  Astronomical Observatory, Chinese Academy of Sciences, 80 Nandan Road, Shanghai 200030, China\\
\email{fgxie@shao.ac.cn}
}

%\date{Submitted} 
\date{}
 
\abstract
  % context heading (optional)
  % {} leave it empty if necessary  
{}
  % aims heading (mandatory)
{We check how the {\it Fermi}/LAT data constrain the physics of hot accretion flows that are most likely present in low-luminosity AGNs.}
  % methods heading (mandatory)
{Using a precise model of emission from hot flows, we studied the flow $\gamma$-ray emission resulting from proton-proton interactions. We explored the dependence of the $\gamma$-ray luminosity on the accretion rate, the black hole spin, the magnetic field strength, the electron heating efficiency, and the particle distribution. Then, we compared the hadronic $\gamma$-ray luminosities predicted by the  model for several nearby Seyfert 1  galaxies with the results of our analysis of 6.4 years of {\it Fermi}/LAT observations of these AGNs.}
  % results heading (mandatory)
{In agreement with previous studies, we find a significant $\gamma$-ray detection in NGC 6814. We were only able to derive upper limits for the remaining objects, although we report marginally significant ($\sim 3 \sigma$) signals at the positions of NGC 4151 and NGC 4258. The derived upper limits for the flux above 1 GeV allow us to constrain the proton acceleration efficiency in flows with heating of electrons dominated by Coulomb interactions, which case is favored by the X-ray spectral properties. In these flows, at most $\sim 10$\% of the accretion power can be used for a relativistic acceleration of protons. Upper limits for the flux below 1 GeV can constrain the magnetic field strength and black hole  spin value; we find these constraints for NGC 7213 and NGC 4151. We also note that the spectral component above $\sim 4$ GeV previously found in the {\it Fermi}/LAT data of Centaurus A\ may be due to hadronic emission from a flow within the above constraint. We rule out this origin of the $\gamma$-ray emission for NGC 6814. For models with a strong magnetohydrodynamic heating of electrons, the hadronic $\gamma$-ray fluxes are below the {\it Fermi}/LAT sensitivity even for the closest AGNs. In these models, nonthermal Compton radiation may dominate in the $\gamma$-ray range if electrons are efficiently accelerated and the acceleration index is hard; for the index $\simeq 2$, the LAT upper limits constrain the fraction of accretion power used for such an acceleration to at most $\sim 5$\%. Finally, we note that the three Seyfert 2 galaxies with high starburst activity NGC 4595, NGC 1068, and Circinus show an interesting correlation of their $\gamma$-ray luminosities with properties of their active nuclei, and we discuss this in the context of the hot flow model.
}
  % conclusions heading (optional), leave it empty if necessary 
{}

 
\titlerunning{$\gamma$-ray activity of Seyfert galaxies}
\authorrunning{Wojaczy\'nski et al.}
\maketitle

\section{Introduction} 
\label{intro} 

Low-luminosity AGNs are probably powered by optically thin, hot accretion flows \citep[for recent reviews see, e.g.,][]{2014SSRv..183...61P,2014ARA&A..52..529Y}. In the central parts of these flows, protons have energies above the threshold for pion production, and then the flows produce considerable $\gamma$-ray fluxes  that are due to the decay of neutral pions. 
This property was noted early in the development of the accretion theory, for instance, by~\citet{1976ApJ...204..187S}, and was
reported for the currently applied  advection-dominated accretion flow (ADAF) models  by \citet{1997ApJ...486..268M}. As estimated in \citet{1976ApJ...204..187S} and \citet{2003MNRAS.340..543O}, the $\gamma$-ray luminosity, $L_{\gamma}$, of a flow surrounding a rapidly rotating black hole may be similar to its X-ray luminosity, $L_{\rm X}$.  However, these studies neglected the general relativistic (GR) transfer of $\gamma$-ray photons as well as the absorption
of these photons in the radiation field of the flow. Improved computations by \cite{2013MNRAS.432.1576N}, hereafter N13, showed that $L_{\gamma}$ is at least an order of magnitude lower than $L_\mathrm{X}$ if these effects are taken into account. The precise $\gamma$-ray emission model, however, has not been directly compared with current observational data.

No prominent $\gamma$-ray signal was found based on observations in the  {\it Fermi}/LAT data from radio-quiet AGNs 
\citep[e.g.,][]{2012ApJ...747..104A,2015arXiv150106054A}, with a few notable exceptions. These include the three X-ray brightest Seyfert 2 galaxies, NGC 4945, NGC 1068, and Circinus \citep{2010A&A...524A..72L, 2012ApJ...755..164A,2013ApJ...779..131H}. However, all three are composite AGN plus starburst systems, and it has not yet been definitively established whether their $\gamma$-ray emission is related with the starburst or with the AGN activity. The upper limits (UL) on the photon flux in the GeV range, derived in \cite{2012ApJ...747..104A}, typically constrain the luminosity ratio to $L_{\gamma}/L_{\rm X} < 0.1$   and in several Seyferts to $L_{\gamma}/L_{\rm X} < 0.01$, which shows that the sensitivity of the {\it Fermi}/LAT surveys has reached the level at which predictions of the hadronic emission from hot accretion flows can be probed. This motivated the detailed comparison we present here.

We analyzed 6.4 years of  {\it Fermi}-LAT data of nearby, low-luminosity AGNs, which means that we use a data set that is more than three years  longer than the one used by \cite{2012ApJ...747..104A}  and is more than two years longer than the recent third catalog of AGNs detected by {\it Fermi}/LAT
\citep{2015arXiv150106054A}. A detailed comparison with the model requires a precisely determined black hole mass and intrinsic X-ray luminosity, therefore we focus here on several of the best-studied objects. As a particularly interesting case, we thoroughly examine the data from NGC 4151, for which the ratio of  $L_{\gamma}/L_{\rm X} < 0.0025$, found in \cite{2012ApJ...747..104A}, is the lowest of the 120 Seyfert galaxies considered in their work.

The physics of particle acceleration and heating in hot flows is not well understood. If non-thermal processes (e.g.,\ magnetic reconnection) take place, a power-law-like distribution of particles may be expected.
We here take into account predictions of hot flow models with proton distributions  that include thermal and nonthermal components. 
We compute the $\gamma$-ray emission from accretion flows following the approach of N13.
We use a GR hydrodynamic description of the flow combined with  GR, Monte Carlo (MC) computations of radiative processes. We improve the model of N13 by refining the hydrodynamic solutions for flows whose proton distribution is dominated by a  nonthermal component.
We obtain solutions for a range of accretion rates that allows comparing our results with objects that have bolometric luminosities of between $\sim 10^{-4}$ and $10^{-2}$ of the Eddington limit. In Sect. \ref{sect:model} we discuss the dependence of the predicted $L_{\gamma}$ on various parameters of the hot flow model, this has not been fully explored in precise computations before. In particular, N13 only considered a weak magnetic field and a fixed accretion rate. In Sect. \ref{sect:obs} we present the results of our analysis of the {\it Fermi}/LAT data together with some results collected from the literature. Our main results are presented in Figs. \ref{fig:1} and \ref{fig:2}. In Sect. \ref{nthel} we briefly consider effects  that are related with the nonthermal acceleration of electrons, in particular, their Compton emission in $\gamma$-rays. In Sect. \ref{sect:discuss} we discuss the
general implications of our results for the physics of hot flows; we also briefly discuss hard spectral states of black hole binaries, where similar hot flows are most likely present.

\section{Model of X- and $\gamma$-ray emission from hot flows}
\label{sect:model}

We applied the model developed in \cite{2012MNRAS.420.1195N,2013MNRAS.432.1576N,2014MNRAS.443.1733N,2015ApJ...799..217N} and \cite{2010MNRAS.403..170X}. Here we briefly summarize its basic properties and some improvements we added here.

We considered a black hole that is characterized by its mass, $M$, and angular momentum, $J$, surrounded by a geometrically thick accretion flow with an accretion rate, $\dot M$. We defined the following dimensionless parameters: $r = R / R_{\rm g}$, $a = J / (c R_{\rm g} M)$, $\dot m = \dot M / \dot M_{\rm Edd}$, where $\dot M_{\rm Edd}= L_{\rm Edd}/c^2$, $R_{\rm g}=GM/c^2$ is the gravitational radius and $L_{\rm Edd} \equiv 4\pi GM m_{\rm p} c/\sigma_{\rm T}$ is the Eddington luminosity.  
We assumed a viscosity parameter, $\alpha=0.3$ and the ratio of the gas pressure (electron and ion) to the magnetic pressure, $\beta$. 

The fraction of the dissipated energy that directly heats electrons is denoted by $\delta$. We considered two limiting cases, one with $\delta=10^{-3}$, for which electron heating  is dominated by Coulomb interactions at higher $\dot m$ and compression at lower $\dot m$, and the other with $\delta=0.5$, for which electron heating is dominated by magnetohydrodynamic (MHD) processes. Our results for $\delta=10^{-3}$ are valid for 
all values of $\delta \la 0.1$.

\begin{figure} 
\centerline{
\includegraphics[width=9cm]{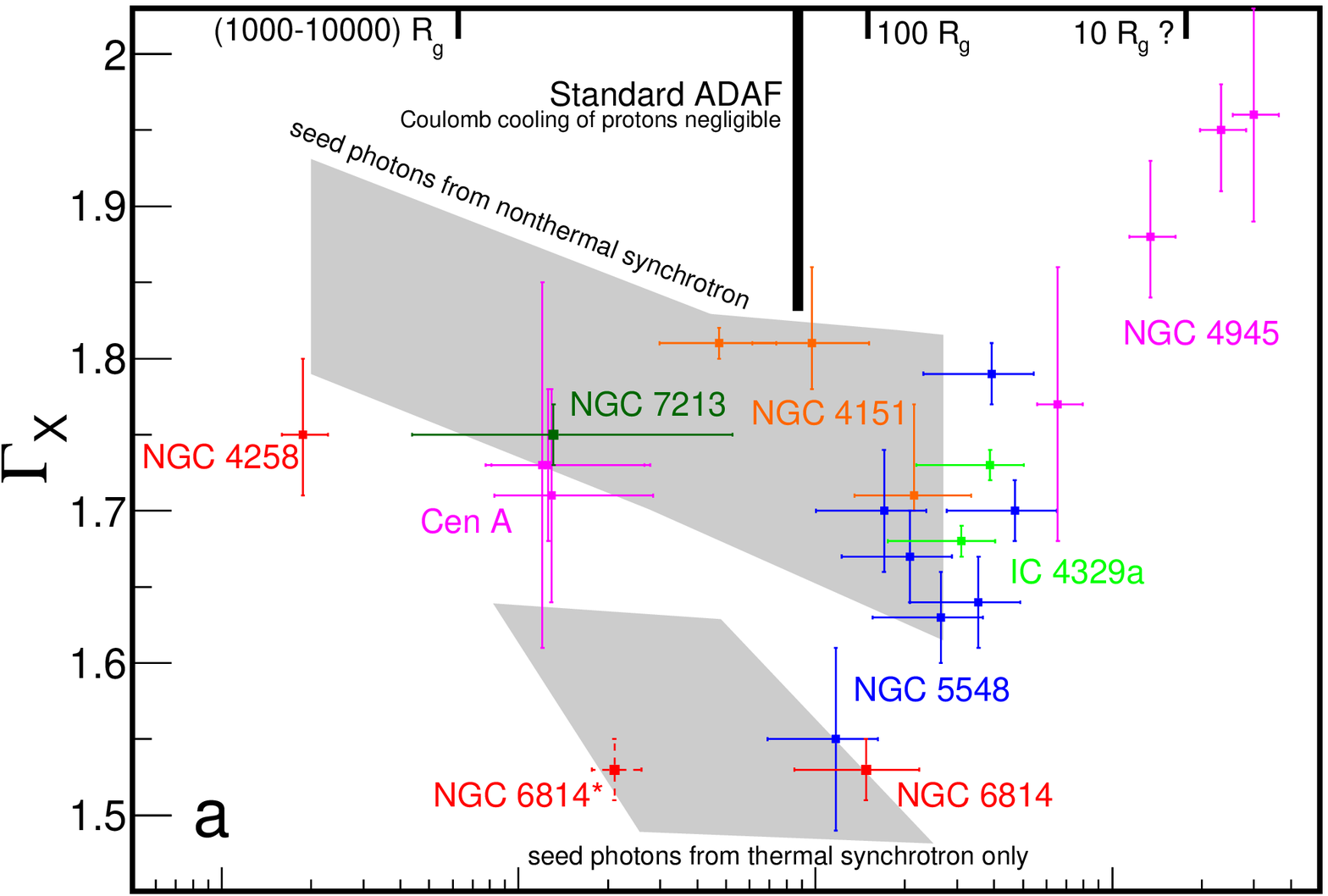}}
\centerline{
\includegraphics[width=9cm]{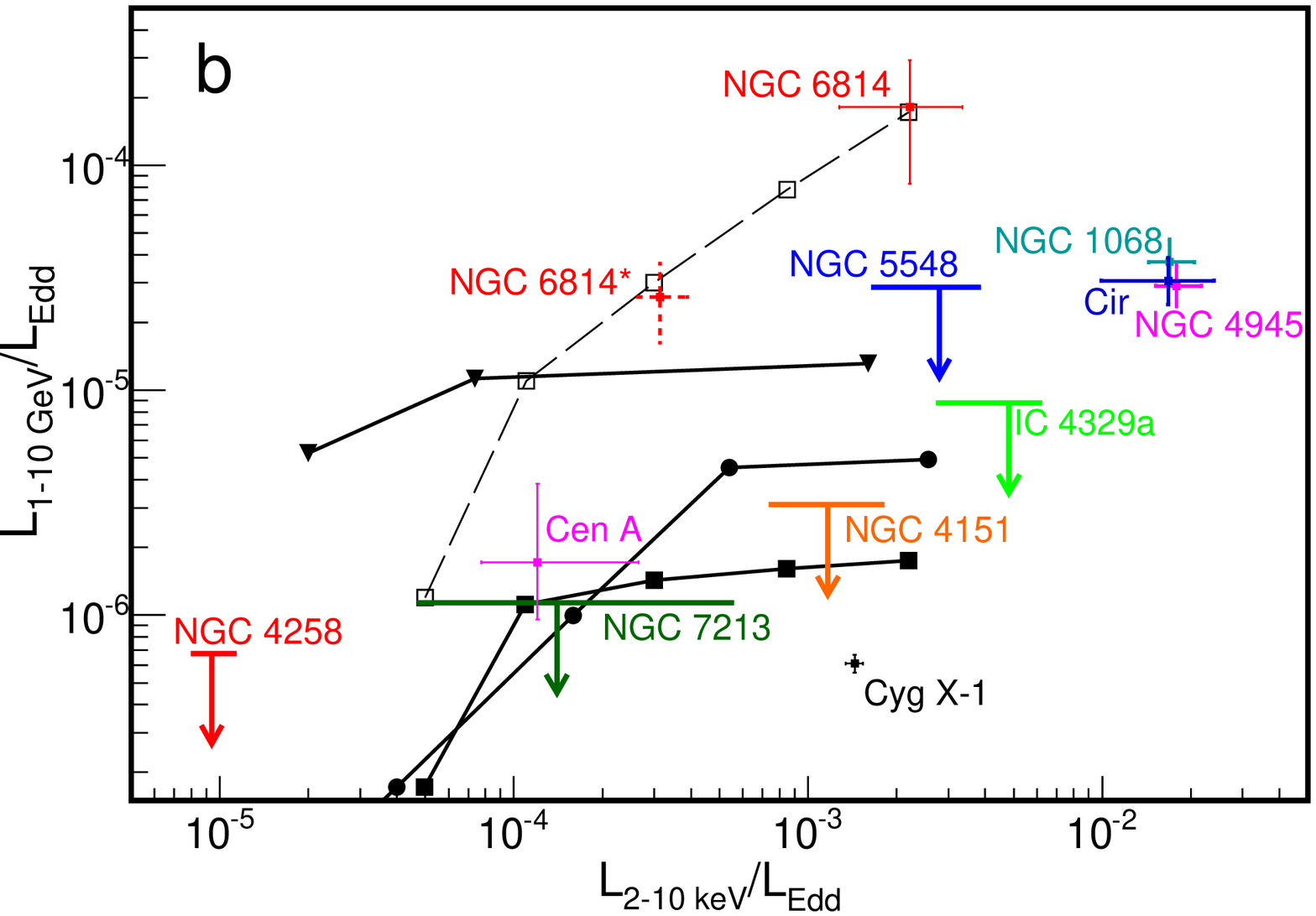}}
\caption{(a) X-ray photon spectral index
as a function of intrinsic $\lambda_\mathrm{2-10\;keV}$. The observational data correspond to the best fits of the high-quality observations described in Sect. \ref{sect:obs}. The gray regions indicate the location of solutions for the model with seed photons from thermal synchrotron only (lower) and including nonthermal synchrotron (of pion-decay or directly accelerated electrons; upper panel). Ticks in the top axis show the approximate value of $r_{\rm tr}$ (see text).
The thick vertical line indicates the maximum luminosity for the standard ADAF solutions (i.e.,\ unaffected by the Coulomb cooling of protons). (b) $\lambda_\mathrm{1-10\;GeV}$ as a function of $\lambda_\mathrm{2-10\;keV}$, see Tables \ref{tab:basic} and \ref{tab:gamma} for the observational data. The solid lines are for model N with $\delta=10^{-3}$, $a=0.95$, $\beta=9$ (triangles), model N with $\delta=0.5$, $a=0$, $\beta=9$ is plotted with circles), model H$_{0.1}$ with $\delta=10^{-3}$, $a=0.95$, $\beta=9$  with
squares, and  all three are the same as in Fig.\ \ref{fig:0}b below. 
The dashed line with open squares denotes the rest-frame 1-10\;GeV luminosity in the last model (i.e.,\  H$_{0.1}$). 
In panel (b), $\lambda_\mathrm{2-10\;keV}$  corresponds to the average X-ray luminosity during the {\it Fermi}/LAT observations (see text). This differs in some objects by up to a factor of 2 relative to panel (a). The point for Cyg X-1 shows the hard state parameters according to \cite{2013MNRAS.434.2380M}.
}
\label{fig:1} 
\end{figure}

The Eddington ratio for a given energy range is denoted by $\lambda$ with
a relevant subscript, for example,\ $\lambda_\mathrm{2-10\;keV}=L_\mathrm{2-10\;keV}/L_{\rm Edd}$
and $\lambda_\mathrm{1-10\;GeV}=L_\mathrm{1-10\;GeV}/L_{\rm Edd}$,
where $L_\mathrm{2-10\;keV}$ and $L_\mathrm{1-10\;GeV}$ are the luminosities in the 2--10 keV and 1--10 GeV
range detected far away from the flow, that is, they are\ affected by GR transfer and $\gamma \gamma$ absorption effects 
(only the dashed line in Fig.\ \ref{fig:1}b shows the rest-frame luminosity).
Similarly, the photon flux integrated over a given energy range is denoted by $F$ 
with a relevant subscript.

We found the global hydrodynamical solutions of the GR structure  equations 
following \cite{2000ApJ...534..734M} with minor improvements as described in \cite{2012MNRAS.420.1195N}.
From this, we found the self-consistent electron temperature
distribution, $T_{\rm e}(r)$, for which the electron energy balance when global Compton cooling is achieved.
This involved several iterations between the solutions of the electron energy
equation  and the MC Comptonization simulations.
The procedure involves the assumption that the flow structure is unaffected 
by changes in $T_{\rm e}$, which constrains the range of exact solutions to $\lambda_\mathrm{2-10\;keV}
\la (1-2) \times 10^{-3}$ (depending on the model, see also \cite{2015MNRAS.447.1692Y}); the maximum  $\lambda_\mathrm{2-10\;keV}$
is denoted by $\lambda_\mathrm{ADAF,max}$ and is indicated by the thick vertical 
line in Fig.\ \ref{fig:1}a.
The Coulomb cooling of protons becomes
strong at higher luminosities, and the flow is characterized by a dramatic dependence on
even small changes of $T_{\rm e}$ (when global Compton cooling is taken into account, cf.\ \cite{2010MNRAS.403..170X}), which  should lead to a collapse of the flow and formation of a standard cold disk at $r > r_{\rm tr}$, with the transition radius, $r_{\rm tr}$, dependent on $\dot m$.

All model results we present here were obtained for flows extending out to $r = 10^4$. 
We currently investigate the regime of $\lambda_\mathrm{2-10\;keV} > \lambda_\mathrm{ADAF,max}$.
Although the results are rather preliminary, we find that (1) up to $\lambda_\mathrm{2-10\;keV} \simeq 3 \times 10^{-3}$ (indicated by the boundary of gray area in Fig.\ \ref{fig:1}a)  the hot flow can occupy the inner region within $r_{\rm tr} \simeq 100$ (this accretion flow geometry is supported by observations, as noted below); (2) up to this $\lambda_\mathrm{2-10\;keV}$,
 the X-ray and the rest-frame $\gamma$-ray emission is correctly approximated by models with flows extending to large $r$,
for which we used the hydrodynamical solution neglecting the Coulomb cooling of protons; and (3) this model (with a flow extending to large $r$) may overestimate the $\gamma$-ray opacity (see below).

\begin{figure*} 
\centerline{
\includegraphics[height=5.5cm]{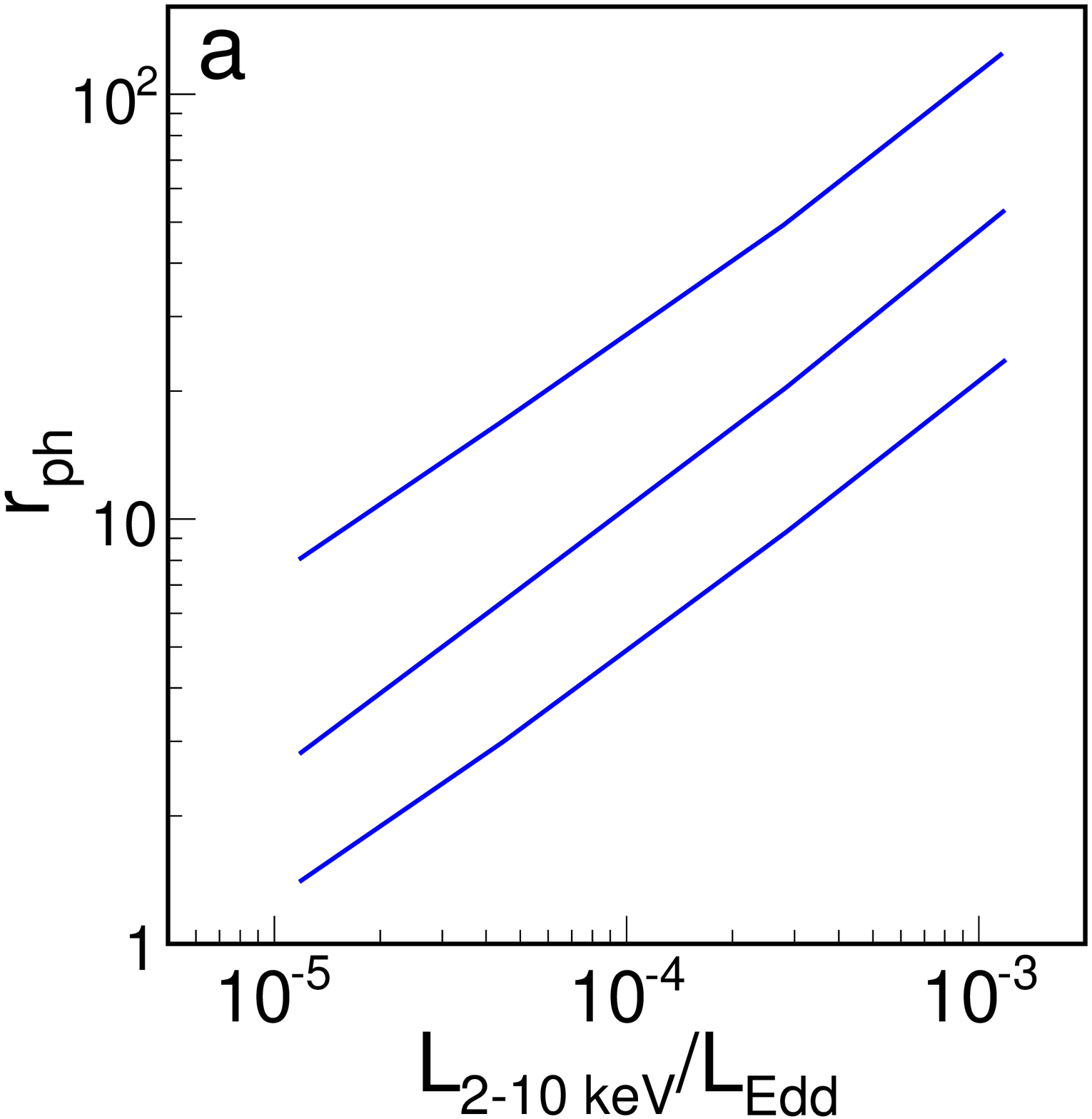}
\includegraphics[height=5.5cm]{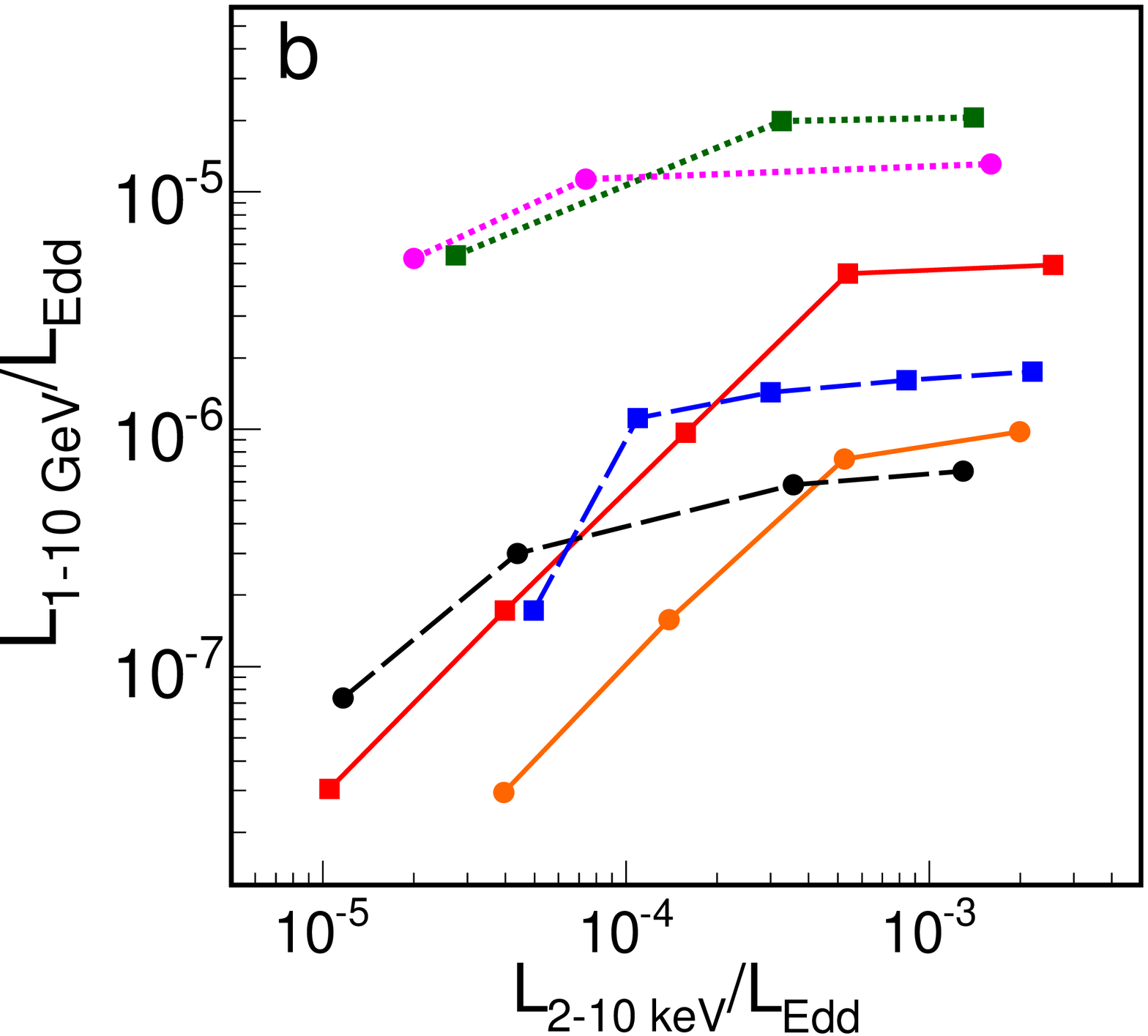}
\includegraphics[height=5.5cm]{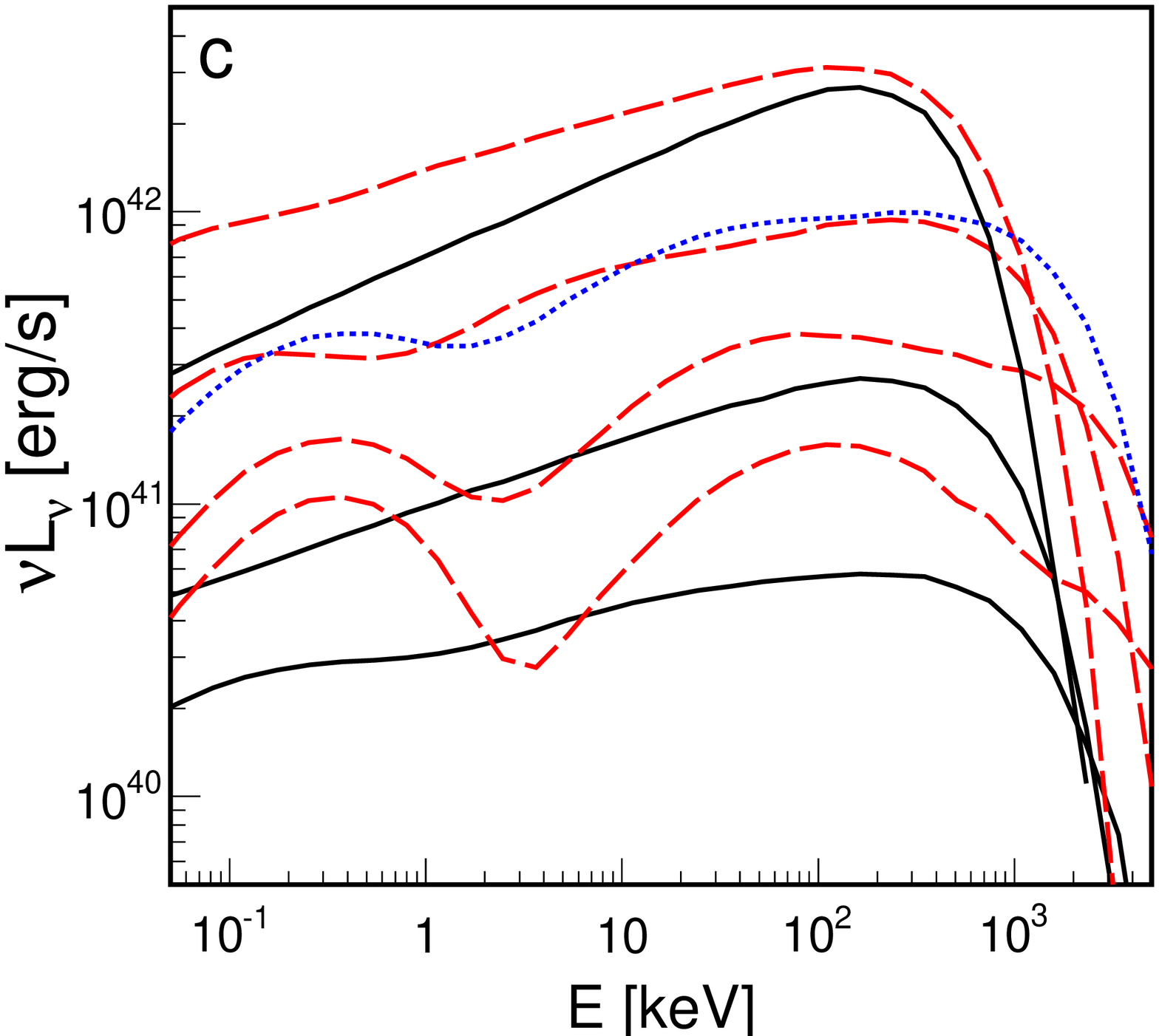}
}
\caption{(a) Photosphere radius for photons with $E=0.1$, 1, and 10 GeV from bottom to top as a function of $\lambda_\mathrm{2-10\;keV}$ in the model with $\delta=10^{-3}$, $a=0.95$, $\beta=1$ (in other models the dependence of $r_{\rm ph}$ on $\lambda_\mathrm{2-10\;keV}$ is very similar); (b) $\lambda_\mathrm{1-10\;GeV}$ as a function of $\lambda_\mathrm{2-10\;keV}$ in models with $\eta_{\rm p} > 0$ and $s=2.1$; the dotted lines plot models N with $\delta=10^{-3}$, (magenta) circles indicate models N with $a=0.95$ and $\beta=9$ with $\dot m=0.05$, 0.1, and 0.3, (green) squares stand for $a=0$, $\beta=9$, $\dot m=0.1$, 0.3 and 0.5; the dashed lines show models H$_{0.1}$  with $\delta=10^{-3}$, (black) circles models with $a=0.95$, $\beta=1$ with $\dot m=0.033$, 0.1, 0.3 and 0.5; (blue) squares indicate $a=0.95$, $\beta=1$ with $\dot m=0.1$, 0.3, 0.5, 0.8 and 1.2;  the solid lines show models N with $\delta=0.5$, (red) squares depict  $a=0$, $\beta=1$ with $\dot m=0.004$, 0.01, 0.033, 0.1 and 0.3; (orange) circles stand for  $a=0.95$, $\beta=9$  with $\dot m=0.003$, 0.01, 0.033 and 0.1. (c) The high-energy part of the observed spectra of leptonic emission. (Black) solid lines indicate model H$_{0.1}$ with $\delta=10^{-3}$, $a=0.95$, $\beta=1$ and $\dot m = 0.033$, 0.1, 0.3; (red) dashed lines show model N with $\delta=0.5$, $a=0$, $\beta=1$ and $\dot m = 0.004$, 0.01, 0.033, 0.1; (blue) dotted line shows model N with $\delta=0.5$, $a=0.95$, $\beta=9$ and $\dot m = 0.01$.}
\label{fig:0} 
\end{figure*}

The minimum luminosity of flows that can be studied with our computational method, determining the lowest $\lambda_{\rm 2-10\;keV}$ (denoted by  $\lambda_{\rm min}$) in all presented models, is
set by the requirement that the energy balance for electrons is determined by radiative cooling (rather than advection). Close to $\lambda_{\rm min}$ , the compressive heating of electrons dominates for $\delta=10^{-3}$, yielding an approximately constant radiative efficiency.
This effect is crucial for assessing the $\gamma$-ray emission in objects with  $\lambda_\mathrm{2-10\;keV} \ll 10^{-4}$ , and we return to this below. 

The magnitude of the hadronic processes involves a significant uncertainty related
to the distribution of proton energies.
At $\lambda_\mathrm{2-10\;keV} < \lambda_\mathrm{ADAF,max}$
the thermalization and cooling timescales for protons are much
longer than the accretion timescale, and the
distribution of protons is determined by poorly understood heating
and acceleration
processes. We  considered three cases:

\noindent
a thermal model T in which all protons have a Maxwellian distribution; 

\noindent
a nonthermal model N, where the total power is
used to accelerate a small fraction of protons, for which we assume  
a power-law distribution, $n_{\rm pl}(\gamma) \propto \gamma^{-s}$ up to $\gamma_{\rm max} = 100$, and 
we assume that the other protons remain cold;  the radius-dependent fraction of the protons
with the power-law distribution is given by Eq. (7) in N13;

\noindent
and a hybrid model H$_{0.1}$ that has 10\%  of
the energy content in nonthermal protons, denoted by $\eta_{\rm p}$,
and 90\% of the energy content in thermal protons; for the nonthermal protons we assume a power-law distribution like that in  model N.

\noindent
The assumptions of these models are
identical to models T, N, and H in N13, except for the energy content of nonthermal component in model H, which was radius-dependent in 
N13, whereas here we assumed a radius-independent $\eta_{\rm p} = 0.1$ (indicated by the subscript in H$_{0.1}$).
For model N, however, we here take into account the effect neglected in N13, namely, a higher compressibility of the flow supported by the pressure of a nonthermal, relativistic proton gas, corresponding to the decrease of the adiabatic index from $\simeq 5/3$ (for a thermal proton gas with typically subrelativistic temperatures) to 4/3. All results for model N in this work 
use the hydrodynamic solutions including this effect. The higher compressibility of the flow 
results in its smaller geometrical thickness and, hence, higher density (see also Fig. 4 in \cite{1998ApJ...504..419P}). For a given set of parameters, the density is higher by a factor of  $\sim 1.5$ 
(for $\beta=1$)  to  $\sim 5$ (for $\beta=9$) in model N than
in models T or H$_{0.1}$.

We note that the dependence of the geometrical
structure on $\beta$ in model N, which was discussed in \cite{2014MNRAS.443.1733N}, is almost completely reduced. The results presented for this model, with both components significantly contributing to the total pressure (i.e.,\ protons and magnetic field), described by the adiabatic index 4/3,  correspond to the maximum density, and hence the maximum Thomson depth, $\tau$, as well as the maximum $L_{\gamma}$ of the flow for a given set of $\dot m$, $a$, $\beta,$ and $\delta$.

To compute the X-ray radiation, we took into account seed photons from nonthermal synchrotron emission, which is crucial to reach 
agreement for the model spectra with the observational data. 
As first noted by \cite{2004MNRAS.354..953Y}, ADAF models that
follow the original formulation by \cite{1995ApJ...452..710N}, that is,\ whose seed photons are provided by thermal synchrotron
emission, predict too high $T_{\rm e}$ and disagree with observational data for AGNs. 
Following the results of  \cite{2015ApJ...799..217N}, we self-consistently included the synchrotron emission 
of pion-decay electrons for $\delta=10^{-3}$, whereas for $\delta=0.5$ we assumed that a fraction $\eta_{\rm e}$ of the electron heating
power was used for their nonthermal acceleration. In the latter case, we assumed $\eta_{\rm e}=0.1$ and a monoenergetic injection of electrons at $\gamma=100$,
for which synchrotron self-absorption is relatively unimportant (in Sect. \ref{sect:accel} we release this assumption and consider the dependence on the electron acceleration distribution). The two gray regions are delineated in Fig.\ \ref{fig:1}a 
to indicate the location of model solutions that include a significant nonthermal synchrotron radiation (upper panel) and solutions with only thermal synchrotron emission (lower panel), see Sect. \ref{sect:discuss} for a more detailed discussion.

\begin{figure} 
\centerline{
\includegraphics[height=5cm]{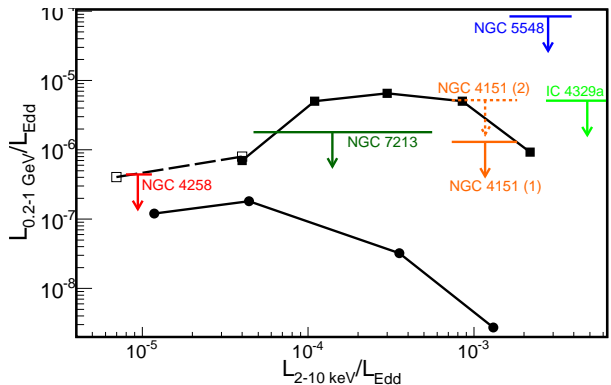}}
\caption{$\lambda_\mathrm{0.2-1\;GeV}$ as a function of $\lambda_\mathrm{2-10\;keV}$, see Tables  \ref{tab:basic}  and  \ref{tab:gamma} for the observational data. All model points are for thermal protons (model T) with $\delta=10^{-3}$; the solid line with circles shows $a=0.95$, $\beta=1$, $\dot m=0.033$, 0.1, 0.3, and 0.5; the solid line with full squares depicts
 $a=0.95$, $\beta=9$, $\dot m=0.1$, 0.3, 0.5, 0.8, and 1.2; the
dashed line with open squares plots the estimate (see text) for an edge-on observer for $a=0.998$, $\beta=9$, $\dot m=0.033,$ and 0.1.
}
\label{fig:2} 
\end{figure}

Using our MC Comptonization code, we  tabulated the angular-, energy-, and location-dependent distribution 
of photons propagating in the central region. Then, to compute $L_{\gamma}$ , we took into account
the absorption of escaping $\gamma$-ray photons on pair creation in interactions with the (tabulated) target photons, strictly following N13. 
In Fig.\ \ref{fig:0}a we show the size, $r_{\rm ph}$, of the $\gamma$-ray photosphere (inside which the flow is  opaque to $\gamma$-rays), which is determined as the radius of the emission point in the equatorial
plane, from which the optical depth along the outward radial trajectory is $\tau_{\gamma \gamma} = 1$. For radial directions close to the symmetry axis, values of $r_{\rm ph}$ determined by a similar condition are lower by a factor of $\sim 1.5$ . Results presented in this work, except for the dashed line in Fig.\ \ref{fig:2}, correspond to spectra averaged over the observation angle, $\theta_{\rm obs}$. A strong dependence on $\theta_{\rm obs}$ only occurs for a rapidly rotating black hole and only at low $\lambda_\mathrm{2-10\;keV}$, at which the contribution from the innermost few $R_{\rm g}$ is not attenuated by $\gamma \gamma$ absorption. In other cases, the difference between $\gamma$-ray fluxes received at different  $\theta_{\rm obs}$ does not exceed a factor of $\sim 2$.
The amount of attenuation due to $\gamma \gamma$ absorption is illustrated by the dashed (rest-frame $L_{\gamma}$) and solid (observed $L_{\gamma}$) lines with squares in Fig.\ \ref{fig:1}b; we note that at high $\lambda_\mathrm{2-10\;keV}$ the internal  $L_{\gamma}$ is higher by two orders of magnitude than observed.

The dependence of  $r_{\rm ph}$ on $\lambda_\mathrm{2-10\;keV}$ is similar in all models (i.e.,\ for different $a$, $\beta,$ or $\delta$). However, the scaling presented in Fig.\ \ref{fig:0}a cannot be simply extrapolated to $\lambda_\mathrm{2-10\;keV} \ga 10^{-3}$, at which this extrapolated  $r_{\rm ph}$ would exceed the flow size of $r_{\rm tr} \simeq 100$. In this case,  the $\gamma$-ray opacity should be comparatively lower, at least in outer parts of the flow (i.e.,\ close to  $r_{\rm tr}$). 
In a flow extending to large $r,$ the opacity for a $\gamma$-ray photon escaping from $r_{\rm em}$ is dominated by head-head interactions with X-ray photons produced at $r >  r_{\rm em}$. For photons escaping from the outer parts of a flow contained within $r_{\rm tr}$, the opacity is dominated by much less efficient head-tail interactions with X-ray photons produced at $r <  r_{\rm em}$.

Below we summarize the dependence of $L_{\gamma}$ on various parameters that is relevant for the comparison presented in Sect. \ref{sect:obs}.
For thermal protons, the bulk of the hadronic radiation is emitted in the 0.1-1 GeV range.
The presence of nonthermal protons is reflected in the production of photons with  $E > 1$ GeV, with a power-law spectrum for a power-law proton distribution.
We consider these two regimes separately.

\subsection{Thermal protons: $\lambda_\mathrm{0.2-1\;GeV}$ vs $\delta$, $\beta,$ and $a$}
\label{sect:t}

The rate of pion production depends on the number of protons with energies above the pion production threshold. For the
thermal distribution of protons, this rate  is extremely sensitive to the proton temperature, $T_{\rm p}$. 
The pion production rate is low for  $T_{\rm p} \la  10^{12}$ K and negligible for $T_{\rm p} \la 4 \times 10^{11}$ K.
The highest $T_{\rm p}$, and accordingly the highest $L_{\gamma}$, can be expected for flows with large $\beta$ and $a$ and small $\delta$. The {\it maximum} $\lambda_\mathrm{0.2-1\;GeV}$ that can be observed for thermal protons, with $\lambda_\mathrm{0.2-1\;GeV} \simeq 10^{-5}$ for $\lambda_\mathrm{2-10\;keV}$ between $\sim 10^{-4}$ and $10^{-3}$, is shown by the line with squares in Fig.\ \ref{fig:2}. In models with $a=0$, $\delta=0.5$ or $\beta=1$, $\lambda_\mathrm{0.2-1\;GeV}$ is smaller at least by a factor of several, and even by orders of magnitude at $\lambda_\mathrm{2-10\;keV} \simeq 10^{-3}$; in all these cases, the model predictions are  below the {\it Fermi} ULs discussed in Sect. \ref{sect:obs}.

The dependence on these three parameters involves the following effects. 
In flows with smaller $\beta$, a larger part of the accretion power is used to build up
the magnetic field strength, therefore, the proton heating power is weaker. As a result, for $\beta=1$ the proton temperature is by a factor of $\sim 2$ lower than for $\beta=9$. 
The related decrease of  $\lambda_\mathrm{0.2-1\;GeV}$  is illustrated in Fig.\ \ref{fig:2}; for $\beta=1$ and $a=0.95$, $T_{\rm p}$ only exceeds $4 \times 10^{11}$ K at $r<6$, hence, at $\lambda_\mathrm{2-10\;keV} > 10^{-4}$ the $\gamma$-ray emission is strongly attenuated by $\gamma \gamma$ absorption.
The increase of $\delta$ yields the increase of electron heating at the expense of the proton heating, and it therefore obviously results in the decrease of the $L_{\gamma}/L_{\rm X}$ ratio.
For the decreasing $a$, a decreasing fraction of the accreted rest-energy is dissipated, therefore, $T_{\rm p}$ decreases.
In the extreme case of $a=0$ and $\beta=1$, $T_{\rm p}$ weakly exceeds $4 \times 10^{11}$ K only close to the event horizon, and no hadronic emission can be expected.

At $\lambda_\mathrm{2-10\;keV} \simeq 10^{-5}$ photons with $E \la 1$ GeV may escape from the ergosphere (the extremely relativistic region located within $r<2$) and some specific effects of the Kerr metric can be observed. These are that gravitational focusing of the photon trajectories toward the equatorial plane enhances the flux received by edge-on observers. Furthermore, the increase of $a$ from 0.95 to 0.998 increases $L_{\gamma}$ by a factor of $\sim 2$; at $r>2$ both values of $a$ give the same $\gamma$-ray production rate. Taking these effects into account, we show the estimated maximum $\lambda_\mathrm{0.2-1\;GeV}$ that can be expected at $\lambda_\mathrm{2-10\;keV} \sim 10^{-5}$ by the dashed line in  Fig.\ \ref{fig:2}. For $\beta=9$ this $\lambda_\mathrm{2-10\;keV}$ is lower than $\lambda_\mathrm{min}$, meaning that\ it corresponds to the regime where advection of electron energies dominates.
This in turn means that the 
lowest $\lambda_\mathrm{2-10\;keV}$ point (for $a=0.998$ and $\dot m = 0.033$) is not based on a precise solution of our model; 
we used the scaling of radiative efficiency, $\propto \dot m^{0.7}$, found in 
\cite{2012MNRAS.427.1580X} for low $\dot m$, to roughly estimate the X-ray luminosity.

\subsection{Nonthermal protons: $\lambda_\mathrm{1-10\;GeV}$ vs $\delta$ and $\eta_{\rm p}$}

The dependence of $\lambda_\mathrm{1-10\;GeV}$ on
$\lambda_\mathrm{2-10\;keV}$ in models with $\eta_{\rm p}>0$  is shown in Fig.\ \ref{fig:0}b.
Here the number of protons above the pion production threshold depends only linearly on the proton heating power. This means
that $L_{\gamma}$ is much less sensitive to both $\beta$ and $a$ than in model T. 
The difference between various models results mostly from different radiative efficiencies (of electrons), which establishes the relation between $\lambda_\mathrm{2-10\;keV}$ and $\dot m$, and accordingly the rest-frame $L_{\gamma} \propto \dot m^2$. The efficiency is determined primarily by the value of  $\delta$. Obviously, $\lambda_\mathrm{1-10\;GeV}$ also depends on $\eta_{\rm p}$. However, we note that the difference between models H$_{0.1}$ and N does not simply involve the difference of $\lambda_\mathrm{1-10\;GeV}$ by a factor of 10, corresponding to an increase of $\eta_{\rm p}$ from 0.1 to 1. The difference is more complex because of the difference between the hydrodynamical solutions that underlie these models.

For $\delta=10^{-3}$, when the heating of electrons is dominated by Coulomb interactions (at larger $\dot m$), the rest-frame $\gamma$-ray luminosity scales roughly linearly with the X-ray luminosity (both $\propto \dot m^2$). For this scaling, combined with the increase of $r_{\rm ph}$ with increasing $\dot m$, the observed $\lambda_\mathrm{1-10\;GeV}$ is roughly constant while $\lambda_\mathrm{2-10\;keV}$ changes by more than an order of magnitude (see dotted and dashes lines in Fig.\ \ref{fig:0}b). At low $\dot m$, when the compressive heating of electrons dominates, leading to $\lambda_\mathrm{2-10\;keV} \propto \dot m$, we note a fast decrease of $\lambda_\mathrm{1-10\;GeV}$ with decreasing $\lambda_\mathrm{2-10\;keV}$; this is most clearly seen for model H$_{0.1}$ with $\beta=9$ (dashed line with squares in Fig.\ \ref{fig:0}b), in which compressive heating dominates for $\dot m \la 0.3$, corresponding to $\lambda_\mathrm{2-10\;keV} \la 10^{-4}$.

In models with $\delta = 0.5$, the radiative efficiencies are higher, therefore, $\lambda_\mathrm{1-10\;GeV}$ is lower (because
of the lower $\dot m$) at a given $\lambda_\mathrm{2-10\;keV}$ than for $\delta=10^{-3}$.
Hence, for $\delta = 0.5$ we only present results for model N; for H$_{0.1}$ the predicted $\gamma$-ray flux is below the sensitivity of LAT. For a constant $\delta$, $\lambda_\mathrm{2-10\;keV} \propto \dot m$ results in a fast decrease of $\lambda_\mathrm{1-10\;GeV}$ with decreasing $\lambda_\mathrm{2-10\;keV}$. Furthermore, for large $\delta$, the radiative efficiency
increases with $a$, as a result,   $\lambda_\mathrm{1-10\;GeV}$ at a given $\lambda_\mathrm{2-10\;keV}$ is by a factor of several larger for $a=0$ than for $a=0.95$ (somewhat counter-intuitively).

The uncertain value of $\delta$ introduces the main uncertainty in testing the hadronic emission model by allowing for a range of $\dot m$ at a given $\lambda_\mathrm{2-10\;keV}$. However, different $\dot m$ also correspond to different Thomson depths of the flow, which is reflected in the shape of the X-ray spectrum. 
The above uncertainty can therefore be reduced by using the X-ray spectral information. 
Figure\ \ref{fig:0}c shows example X-ray spectra for large and small $\delta$. Spectra of  large-$\delta$
models in general deviate from a power-law shape at $\lambda_\mathrm{2-10\;keV} \la 10^{-4}$, that is,\ for $\dot m \la 0.01$ with $a=0.95$ and $\dot m \la 0.03$ with $a=0$.
For low-$\delta$ models, deviations from a power-law shape are weak down to $\lambda_\mathrm{2-10\;keV} \sim 10^{-5}$ (the bottom solid line in Fig.\ \ref{fig:0}c).

We considered various proton distribution slopes between $s=2.1$ and 2.7.
In all models, the lowest $\lambda_\mathrm{1-10\;GeV}$ corresponds to $s=2.1$ (for $s=2.6$, $\lambda_\mathrm{1-10\;GeV}$ is larger
by up to a factor 2), and conservatively we assumed this $s$ for Fig.\ \ref{fig:1}b.

\begin{table}

\begin{center}

\caption{Adopted 2-10 keV Eddington ratio, black hole mass, distance,
and references for the data. The references are given in the
order X-ray
measurement, $M,$ and (if available) distance.}

\begin{tabular}{lcccl}

\hline

 Source &  $\lambda_{2-10\;{\rm keV}}$  &  $M$ & $D$  & Refs. \\

                  &  $\times 10^{-2}$                 & $10^{7}M_{\odot}$
&   (Mpc)   &  \\

\hline

IC 4329a          & $0.48$ & $13^{+10}_{-3}$ & $68.4$  &
1\tablefootmark{a},2  \\

NGC 4151        & $0.12$ & $5.4 \pm 1.8$ & $19.0$       &
3\tablefootmark{b},4,4 \\

NGC 4258       & $0.00094$  & $3.6$ & $7.2$ & 5\tablefootmark{c},6,7  \\

NGC 5548       & $0.28$ &   $3.2^{+2.3}_{-0.9} $ &$72.7$        & 8,9 \\

NGC 6814       & $0.15$ & $0.26^{+0.19}_{-0.09}$& $22.8$             &
10\tablefootmark{d},9,11  \\

NGC 6814$^*$       & $0.021$ & $1.9 \pm 0.4$ & $22.8$             &
10\tablefootmark{d},12,11  \\

NGC 7213       & 0.014 & $8.0^{+16.0}_{-6.0}$ & $22.0$  & 13,14,11  \\

Cent.~A         & $ 0.012$ & $5.5 \pm 3.0$ & $3.8$ & 15,16,17  \\

Circinus         & 1.1--2.3  & $0.17 \pm 0.03$ & $4.2$  & 18,19,11 \\

NGC 1068      & $ 1.7$ &   $1.0$    & $14.4$ & 20,21,11  \\

NGC 4945      & $ 1.8 $ &  $0.14$  & $3.6$   & 22,23,24  \\

\hline

\end{tabular}

\tablefoot{$\lambda_{2-10\;{\rm keV}}$ given in the second column involves
rescaling (based on the {\it Swift}/BAT data, see text) of the average
intrinsic luminosity given in the reference by a factor of
\tablefoottext{a}{1.3;}\tablefoottext{b}{1.2;}\tablefoottext{c}{2;}
\tablefoottext{d}{1.5.}}

\tablebib{(1)~\cite{2014ApJ...788...61B}, (2)~\cite{2009ApJ...698.1740M},
(3)~\cite{2010MNRAS.408.1851L}, (4)~\cite{2014Natur.515..528H},
(5)~\cite{2009ApJ...691.1159R}, (6)~\cite{1995Natur.373..127M},
(7)~\cite{1999Natur.400..539H}, (8)~\cite{2012ApJ...744...13B},
(9)~\cite{2015MNRAS.448.3070P}, (10)~\cite{2013ApJ...777L..23W},
(11)~\cite{1988ngc..book.....T}, (12)~\cite{2009ApJ...705..199B},
(13)~\cite{2010MNRAS.408..551L}, (14)~\cite{2014MNRAS.438.3322S},
(15)~\cite{2011ApJ...743..124F}, (16)~\cite{2009MNRAS.394..660C},
(17)~\cite{2010PASA...27..457H}, (18)~\cite{2014ApJ...791...81A},
(19)~\cite{2003ApJ...590..162G}, (20)~\cite{2014arXiv1411.0670B},
(21)~\cite{1996ApJ...472L..21G}, (22)~\cite{2014ApJ...793...26P},
(23)~\cite{1997ApJ...481L..23G}, (24)~\cite{2009AJ....138..323T}}

\label{tab:basic}

\end{center}

\end{table}

\begin{table}

\begin{center}

\caption{Integrated photon flux, $F$, and the photon index, $\Gamma$, of
power-law fits, or 95\% confidence level ULs with the assumed $\Gamma$,
and the corresponding Eddington ratio for 1--10 GeV (upper part) and
0.2--1 GeV (lower part).}

\begin{tabular}{lccc}

\hline

Source & $\Gamma$ &                         $F$ & $\lambda_{\gamma}$  \\

       &          & $10^{-10}\;{\rm ph/cm^2/s}$ & $\times 10^{-5}$    \\

\hline

\hline

 \multicolumn{4}{c}{Energy range: 1--10 GeV }

\\

\hline

IC 4329a &  2.7 & $<\;0.46$ & $<\;0.49$   \\

IC 4329a &  2.1 & $<\;0.68$ & $<\;0.88$   \\

NGC 4151 \tablefootmark{a} & 2.7 & $<\;1.2$ & $<\;0.23$   \\

NGC 4151\tablefootmark{a} & 2.1 & $<\;1.3$ & $<\;0.31$   \\

NGC 4151\tablefootmark{b} & 2.1 & $<\;1.8$ & $<\;0.43$   \\

NGC 4151\tablefootmark{c} & 2.1 & $<\;1.8$ & $<\;0.43$   \\

NGC 4151\tablefootmark{d} & 2.1 & $<\;2.0$ & $<\;0.48$   \\

NGC 4258    & 2.7 & $<\;1.1$ & $<\;0.043$   \\

NGC 4258   & 2.1 & $<\;1.5$ & $<\;0.067$   \\

NGC 5548   & 2.1 & $<\;0.3$ & $<\;1.8$   \\

NGC 5548   & 2.7 & $<\;0.6$ & $<\;2.9$   \\

NGC 6814   &$2.6 \pm 0.1$ & $3.0 \pm 0.7$ & $18 \pm 10$   \\

NGC 6814$^*$   &$2.6 \pm 0.1$ & $3.0 \pm 0.7$ & $2.6 \pm 1$   \\

NGC 7213   &$2.7$ & $<\;0.4$ & $<\;0.07$   \\

NGC 7213   &$2.1$ & $<\;0.5$ & $<\;0.11$   \\

Cent.~A$^\mathrm{(1)}$ & $2.1 \pm 0.2$ & $18.0 \pm 2.7$ & $
0.17^{+0.2}_{-0.1}$  \\

Circinus$^\mathrm{(2)}$ & $2.4 \pm 0.1$ & $9.2 \pm 1.4$ & $ 3.1\pm 0.7$ \\

NGC 1068$^\mathrm{(2)}$ & $2.3 \pm 0.1$ & $5.4 \pm 0.8$ & $ 3.7 \pm 0.8$  \\

NGC 4945$^\mathrm{(2)}$ & $2.4 \pm 0.1$ & $9.8 \pm 1.2$ & $ 2.9 \pm 0.6$  \\

\hline

 \multicolumn{4}{c}{Energy range: 0.2--1  GeV } \\

\hline

IC 4329a &  4.0 & $<\;3.3$ & $<\; 0.51$   \\

NGC 4151 (1)\tablefootmark{a} & 4.0 & $<\;4.5$ & $<\;  0.13$   \\

NGC 4151 (2)\tablefootmark{b} & 4.0 & $<\;20$ & $<\; 0.52$   \\

NGC 4258 & 4.0 & $<\;7.7$ & $<\; 0.044$   \\

NGC 5548 & 4.0 & $<\;12$ & $<\; 8.4$   \\

NGC 7213 &$4.0$ & $<\;7.1$ & $<\; 0.18$   \\

\hline

\end{tabular}

\tablefoot{\tablefoottext{a}{using the full data set of 6.4 years and
including source \texttt{S} in the model;} \tablefoottext{b}{$\sim
4.9$-year data set (6.4 years without period S), source \texttt{S}
not included;} \tablefoottext{c}{data from the first three years, \texttt{S}
included in the model;} \tablefoottext{d}{data from the first three years,
\texttt{S} not included in the model.}}

\tablebib{

(1)~\cite{2013ApJ...770L...6S}, parameters of the 'second' component above
4 GeV; (2)~\cite{2015ApJS..218...23A}.}

\label{tab:gamma}

\end{center}

\end{table}

\section{Sample and data analysis}

\label{sect:obs}

We considered nearby AGNs with spectral properties consistent with the hot flow scenario. 
This includes several Seyfert 1 galaxies: \object{NGC 4151}, \object{NGC 5548}, \object{IC 4329a}, \object{NGC 6814}, \object{NGC 4258}, and \object{NGC 7213}. We also took into account the FR I galaxy \object{Centaurus A}, whose X-ray radiation may be dominated by a Seyfert-like emission. Spectral parameters of the intrinsic X-ray emission, measured in all these objects, agree with our model of thermal Comptonization in hot flows (see, e.g.,\ Fig \ref{fig:1}a). Furthermore, they show direct evidence of the lack of an optically thick disk in the central region. For objects with the lowest luminosities (NGC 4258 and NGC 7213) studies of the Fe K$\alpha$ line place the inner edge of optically thick disk between $r_{\rm tr} \sim 10^3$ and $\sim 10^4$ \citep{2009ApJ...691.1159R,2010MNRAS.408..551L}.
In more luminous objects (NGC 4151, NGC 5548, IC 4329a, and NGC 6814), the measured Fe K$\alpha$ line widths, $\la 100$ eV
\citep{2010MNRAS.408.1851L,2013ApJ...777L..23W,2012ApJ...744...13B,2014ApJ...788...61B}, indicate truncation of the disk at
$r_{\rm tr} \ga 100$; a similar location of an optically thick disk is indicated by reverberation measurements 
\citep{2015ApJ...806..129E}. These accretion geometries are consistent with the model of
spectral evolution developed for black hole binaries, for instance,\ by \cite{1997ApJ...489..865E},
with  $r_{\rm tr}$ changing in response to changes in $\dot m$ \citep[for a review see][]{2007A&ARv..15....1D}.

We also considered three Seyfert 2 galaxies, \object{NGC 1068}, \object{Circinus}, and \object{NGC 4945}. All three are observed to have high bolometric luminosities, above the range that can be precisely studied with
the current version of our model.
Nevertheless, comparing their $\gamma$-ray luminosities with properties of their active nuclei
appears to be interesting.

For most AGNs in our sample, direct measurements of the masses of their supermassive black holes are available, only for IC 4329a and NGC 7213 did we use estimates based on the velocity dispersion. The  adopted  $M$  and distance values are given in Table \ref{tab:basic}. For NGC 5548 and IC 4329a we assumed the luminosity distances for the following cosmological
parameters: $\Omega_{\rm m} = 0.27$, $\Omega_\Lambda = 0.73$, and $H_0 = 71$ km s$^{-1}$ Mpc$^{-1}$. 
The error bars on the values of $\lambda$ in Figs. \ref{fig:1} and \ref{fig:2} are mostly due to uncertainties in the determination of $M$.
Of the measurements of $M$ that are based on H$_2$0 megamaser kinematics, the uncertainty is reported only for Circinus; for the remaining cases (NGC 1068, NGC 4258, and NGC 4945) we assumed the same relative uncertainties in $M$ for Figs. \ref{fig:1} and \ref{fig:2}. For NGC 6814 the difference of measurements of $M$ between \cite{2009ApJ...705..199B} and 
\cite{2015MNRAS.448.3070P} amount to an order of magnitude, so we present results for both values; the case with the higher value of $M$ is denoted by a star superscript in Tables \ref{tab:basic} and \ref{tab:gamma} and Fig.\ \ref{fig:1}.

\begin{figure} 
\centerline{
\includegraphics[width=9cm]{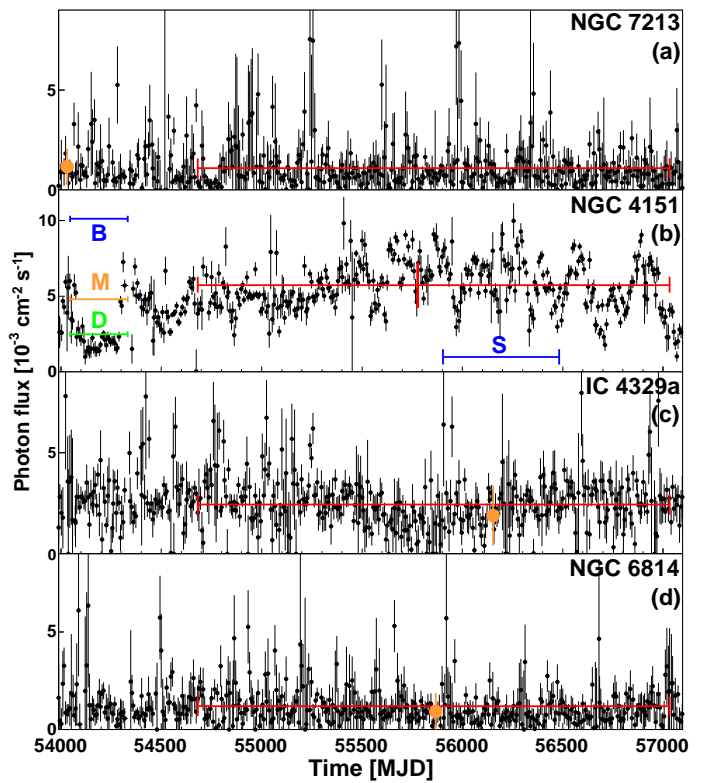}}
\caption{Small black circles with error bars show {\it Swift}/BAT 15-50 keV light curves in seven-day bins for NGC 7213, NGC 4151, IC 4329a, and NGC 6814. The red vertical line in each panel shows the average flux during the 6.4 years of analyzed {\it Fermi}/LAT observations. The large orange circles in panels (acd) show the times of the adopted {\it Suzaku} and {\it NuStar} observations and the average flux in time bins equal to the duration of these pointing observations. The red vertical line in panel (b) indicates the end of the three-year period used for comparison with \cite{2012ApJ...747..104A}. The blue line in the bottom of panel (b) delineates period S (see Sect. \ref{sect:4151}). The flux levels of dim (D), medium (M), and bright (B) states, defined in \cite{2010MNRAS.408.1851L}, are shown in the left part of panel (b).}
\label{fig:lc} 
\end{figure}

Table \ref{tab:gamma} shows the results of our analysis of  the {\it Fermi}/LAT data, described below, except for NGC 1068, Circinus, NGC 4945, and Centaurus A, for which  we used the results of previous studies collected from literature. In Figs. \ref{fig:1}b and \ref{fig:2} we show the measured $\lambda_\mathrm{0.2-1\;GeV}$ and $\lambda_\mathrm{1-10\;GeV}$, or UL on them, as a function of the intrinsic $\lambda_\mathrm{2-10\;keV}$.  
To find the latter, we used results of detailed X-ray spectral studies (specified below) of high-quality data from {\it Suzaku}, {\it NuStar,} or simultaneous XMM and INTEGRAL observations, which allow distinguishing the primary X-ray emission and reflection
or absorption components. The parameters of these fits are shown in Fig.\ \ref{fig:1}a. Using the {\it Swift}/BAT \citep{2013ApJS..209...14K} light curves\footnote{http://swift.gsfc.nasa.gov/results/transients} , we find that even for the most variable objects (NGC 7213, IC 4329a, and NGC 6814; their light curves are shown in Fig.\ \ref{fig:lc}) the adopted X-ray observations correspond to the flux levels 
close to the average {\it Swift}/BAT flux during the analyzed 6.4 years of {\it Fermi}/LAT observations;
in some objects we used the intrinsic   $\lambda_\mathrm{2-10\;keV}$ rescaled to account for a small difference, see Table \ref{tab:basic}.

\subsection{LAT data analysis}

For NGC 4151, NGC 5548, NGC 4258, NGC 6814, NGC 7213, and IC 4329a we analyzed  6.4 years of   the {\it Fermi}/LAT   data, comprising observations carried out between 2008 August 4 and 2015 January 10. For each object, events were selected from a region with a radius of $15^\circ$ centered on the position of the analyzed source. We performed the unbinned likelihood analysis using the v9r33p0 {\it Fermi} Science Tools with CALDB instrument response functions.  We used the standard templates for the Galactic (\texttt{gll\_iem\_v05\_rev1.fits}) and the isotropic (\texttt{iso\_source\_v05\_rev1.txt}) backgrounds. 
In the likelihood analysis we took into account all sources reported in the {\it Fermi}/LAT Third Source Catalog \citep[][hereafter 3FGL]{2015ApJS..218...23A} within a radius of 15$^\circ$ around the analyzed object.
Each of the catalog sources was modeled with a best-fit spectral function (as specified in the catalog, in most cases a power-law)
with parameters left free in the model fitting. 

Except for NGC 6814 we did not find statistically significant signals. We 
derived the 95\% confidence level ULs for the photon flux using the Bayesian method.

We first checked how the use of both an extended data set and improved models (based on 3FGL) of the celestial regions of interest 
affects the UL values. Assuming the $\gamma$-ray photon spectral index $\Gamma=2.5$, 
we found ULs for $F_\mathrm{>0.1\;GeV}$ lower by a factor $\sim 4$ (for NGC 7213 and IC 4329a) and $\sim 2$ 
(for NGC 5548) than the corresponding UL in \citet{2012ApJ...747..104A}. For NGC 4151 the UL value is somewhat dependent on the approach to modeling (as discussed below), but in general, we were unable to reduce it significantly below the value quoted by \citet{2012ApJ...747..104A}.

\begin{figure*}[t]
\centerline{\includegraphics[width=95mm]{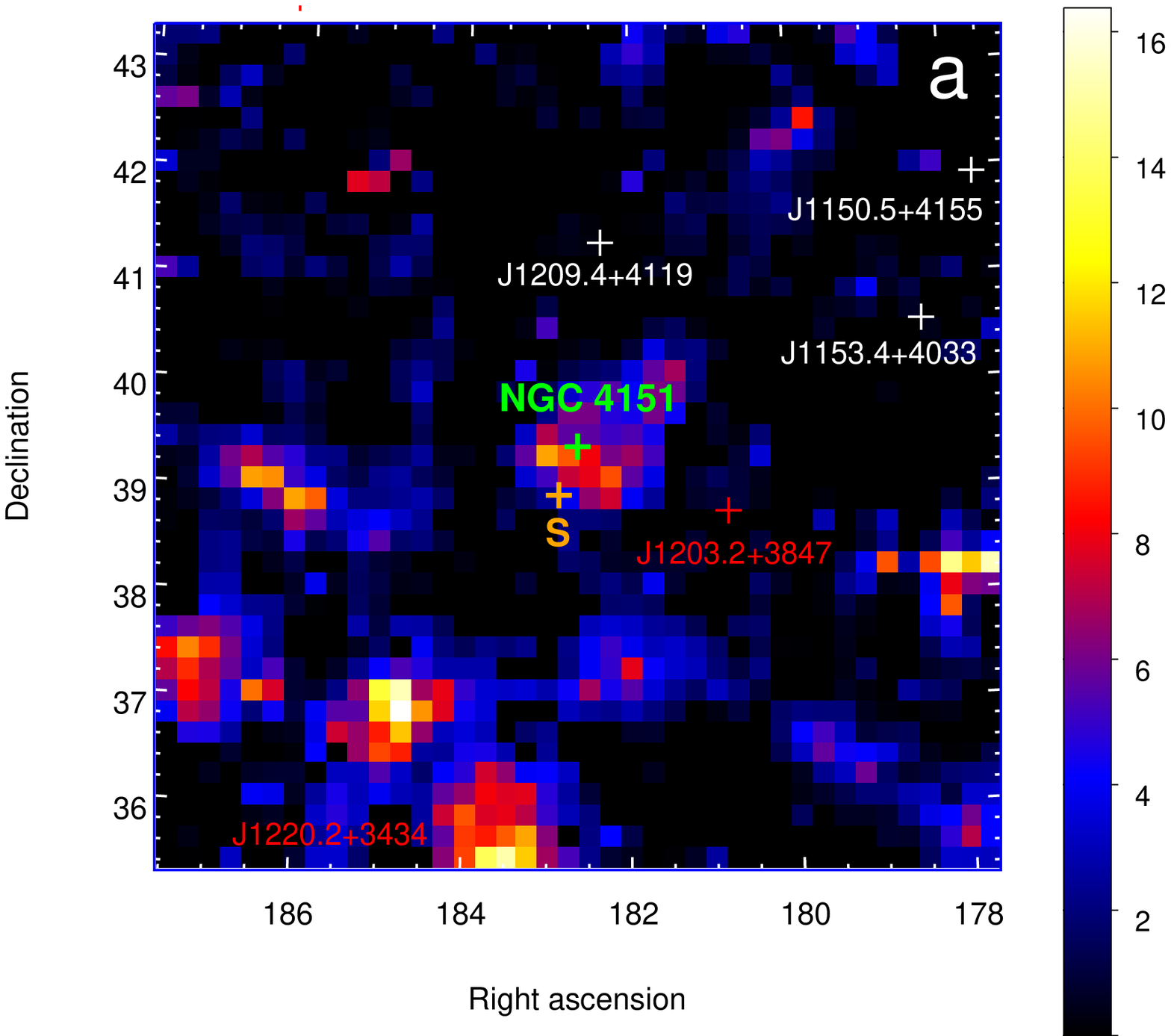} \includegraphics[width=95mm]{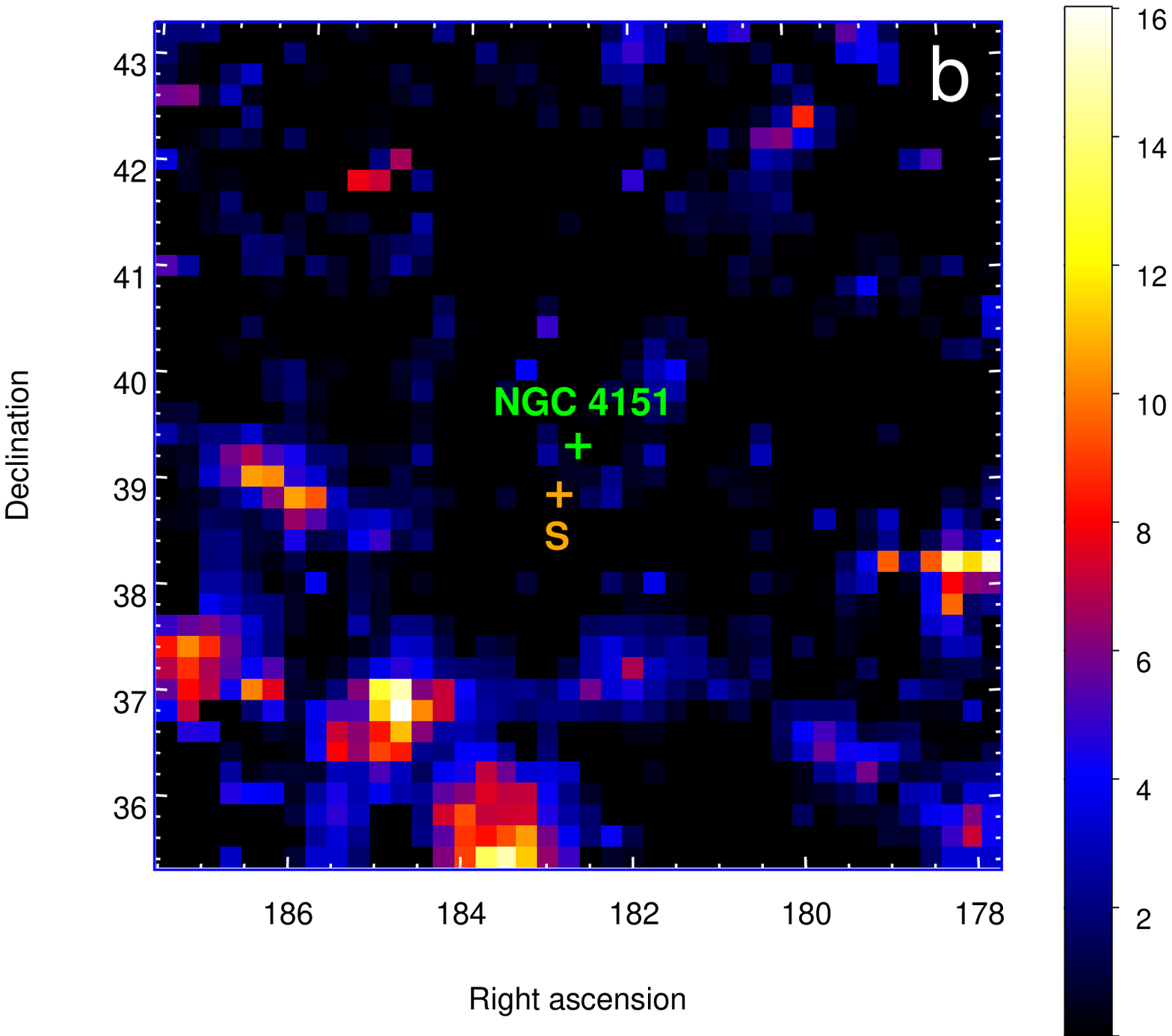}}
\centerline{\includegraphics[width=95mm]{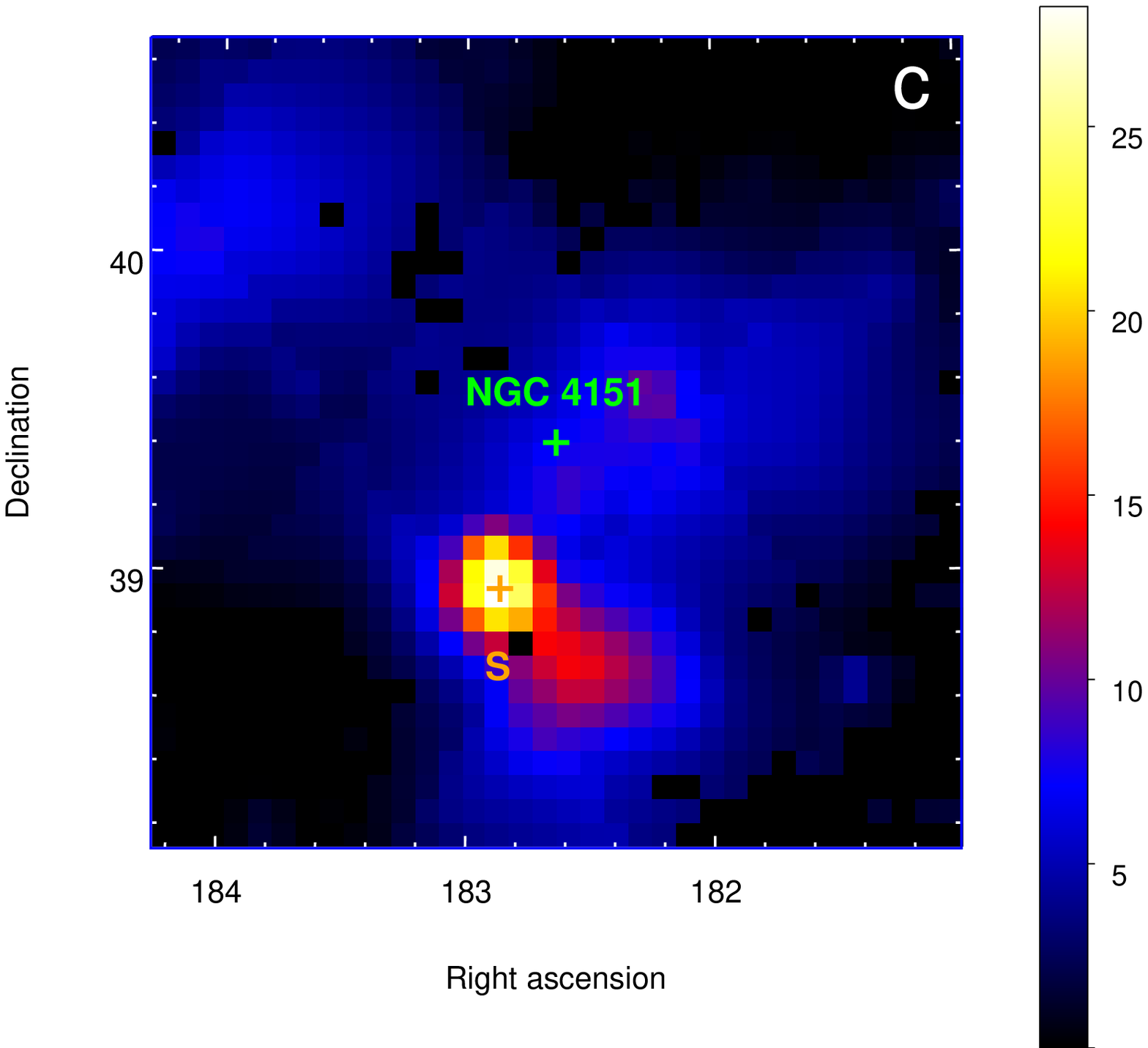} \includegraphics[width=95mm]{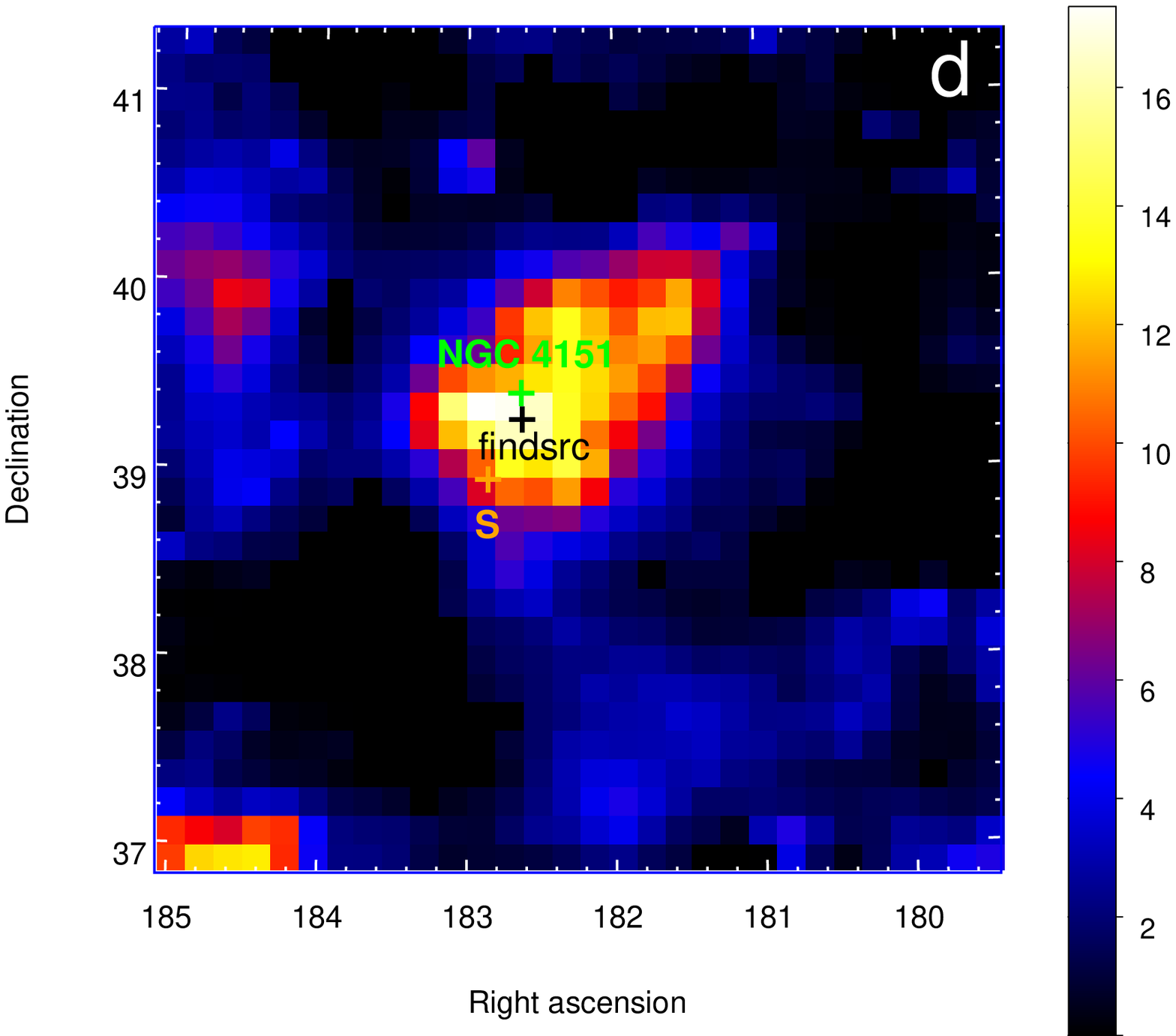}}
\caption{TS maps for the region around NGC 4151; in all panels the green cross shows the location of NGC 4151 and  the orange cross the location of source \texttt{S}. In all panels only the 3FGL sources were  subtracted from the maps, except for panel (b), where an additional point source was added at the position of NGC 4151. (a) The $8\degr \times 8\degr$ degrees region with a pixel size of $0\fdg2$. The map was built for the energy range 0.2 -- 30 GeV, neglecting period S (see text);
the red and white crosses indicate the location of 3FGL sources. 
(b) The same as in (a), but the model includes a point source at the position of NGC 4151 to compensate for the residual seen in (a). 
(c) The $2\fdg55 \times 2\fdg55$ degree region with a pixel size of 0\fdg075  for 0.2-30 GeV for period S.
(d) The $4\fdg8 \times 4\fdg8$ region with a pixel size of 0\fdg075 for 0.2 - 6 GeV and the full period of 6.4 years; the black cross shows the location of a point source found with \texttt{gtfindsrc} under the assumption that the residual seen in the map is produced by a single source; source \texttt{S} is not seen in this map because photons with $E>10$ GeV are excluded.} 
\label{fig:tsmap}
\end{figure*}

The pion decay spectra can only be described by a simple power-law in limited energy ranges.
For the thermal distribution of protons, the $\pi^0$-decay spectrum can be very roughly approximated by a very soft power-law, with $\Gamma \simeq 4$, above $\simeq 0.2$ GeV. For the power-law proton distribution, it is a power-law above $\sim 1$ GeV.
Then, to compare with predictions of model T, we assumed $\Gamma=4$ and found UL for $F_\mathrm{0.2-1\;GeV}$.
To compare with predictions of models with $\eta_{\rm p} > 0$, we found UL for $F_\mathrm{1-10\;GeV}$
assuming different values of $\Gamma$ between 2.1 and 2.7. For all AGNs, except for NGC 5548, the highest ULs correspond to 
$\Gamma = 2.1$ (the difference for other values of $\Gamma$ does not exceed a factor of 2), and we used them in Fig.\ \ref{fig:1}b.
The comparison in Fig.\ \ref{fig:1}b is presented in the most conservative manner, meaning that\ we used the lowest $\lambda_\mathrm{1-10\;GeV}$ predicted by the model (for $s=2.1$) and the highest UL value (for $\Gamma = 2.1$).

\subsection{NGC 4151}
\label{sect:4151}

At the position of NGC 4151 we found a signal with the test-statistic significance TS $\simeq 17$ for $E=0.2-6$ GeV, 
see Fig.\ \ref{fig:tsmap}d. It most likely contains a contribution from a new $\gamma$-ray source found in our analysis, denoted below by \texttt{S}, shifted by only $\sim 0\fdg5$ from NGC 4151. 
Interestingly, however, the \texttt{gtfindsrc} best-fit location for this signal (shown by the black cross), assuming that it comes from a single source, is shifted from the position of NGC 4151 by only 
8\arcmin. Source \texttt{S} is strongly variable, which hinders a proper assessment of its contribution to the $\gamma$-ray signal. The details of our analysis are as follows.

We first examined the significances of $\gamma$-ray signals around NGC 4151 by means of their TS values. This revealed three new sources in the region, which were not reported in the {\it Fermi}/LAT Second Source Catalog  and hence not included in the model used in \citet{2012ApJ...747..104A}.
For each of these sources we used  the  \texttt{gtlike} and \texttt{gtfindsrc} tools to find their significance,  best-fit position, and spectral parameters. Two of them (indicated by the red symbols in Fig.\ \ref{fig:tsmap}a) have recently been reported in 3FGL, with parameters approximately consistent with those estimated in our analysis. They are 
3FGL J1220.2+3434, shifted from NGC 4151 by $\sim 5\degr$, for which our analysis gives $TS \simeq 164$, $\Gamma \simeq 2.2 \pm 0.1$ and $F_\mathrm{>0.1\;GeV}  \simeq (8.4 \pm 1.5) \times 10^{-9}$ ph s$^{-1}$ cm$^{-2}$, and
3FGL J1203.2+3847, shifted from NGC 4151 by  $\sim 1.5$\degr, for which we derived $TS \simeq 32$,  $\Gamma \simeq 2.3 \pm 0.2$ and $F_\mathrm{>0.1\;GeV}  \simeq (3.5 \pm 1.4) \times 10^{-9}$ ph s$^{-1}$ cm$^{-2}$.
These sources are included in the model used for all results presented in this work.

The position of the third source, \texttt{S}, $\sim 0\fdg5$ from NGC 4151,
is determined by several photons with energies between 10 and 20 GeV,
which arrived from the same direction (within $\sim$8\arcmin) between December 2011 and June 2013;
we assume that the source exhibited an outburst activity during this 1.5-year period, which we denote as period S.
Source \texttt{S} is clearly seen in the TS map built for period S alone, see  Fig.\ \ref{fig:tsmap}c, whereas the map built without this period only shows a weak residual, centered on NGC 4151 and not on \texttt{S}, see  Fig.\ \ref{fig:tsmap}a. After subtracting all 3FGL sources, we find for source \texttt{S} (1) TS $\simeq 30$, $\Gamma = 1.78 \pm 0.37$, $F_\mathrm{>0.1\;GeV} \simeq 1.2 \times 10^{-9}$ ph s$^{-1}$ cm$^{-2}$ using the data from period S, and (2)   TS $\simeq 22$, $\Gamma = 2.17 \pm 0.20$, $F_\mathrm{>0.1\;GeV} \simeq 2.1 \times 10^{-9}$ ph s$^{-1}$ cm$^{-2}$ using the full 6.4-year data set. 
Source \texttt{S} is not reported in 3FGL, which was built using data up to 2012 July, that i,\ covering only $\sim 30\%$ of period S.
For the data from period S, \texttt{gtfindsrc} gives the location of \texttt{S} at $\alpha_{J2000} = 12^\mathrm{h} 11^\mathrm{m} 27^\mathrm{s}$, $\delta_{J2000} = 38\degr 56\arcmin 48\arcsec$.   
A possible candidate for this source is a BL Lac object,
2E 1209.0+3917, only $\simeq 4\arcmin$ from this location.   

For a 4.9-year data set, without the period S, \texttt{gtlike} shows no signal (TS $\simeq 1.5$) at the \texttt{S} position.
The residual seen in the TS map for this data set, Fig.\  \ref{fig:tsmap}(a),  can be fully compensated for by adding the point source at the position of NGC 4151 (compare with panel b), and \texttt{gtlike}  
gives $TS \simeq 8$, $\Gamma = 2.7 \pm 0.3$ and $F_\mathrm{>0.1\;GeV} = (1.5 \pm 0.6) \times 10^{-9}$ for this source. 

The presence of a variable source \texttt{S} results in an ambiguity for the UL on the $\gamma$-ray flux from NGC 4151, especially below 1 GeV, where the point spread function of LAT is much higher than the angular separation between   \texttt{S} and NGC 4151.
In our analysis we considered two variants:

\noindent
(1) assuming that \texttt{S} strongly softened before and after period S, but still provided some contribution at low energies, we used the full 6.4-year data set and included source \texttt{S} in the model in addition to the 3FGL sources, which yielded the UL values denoted by superscript 'a' in Table \ref{tab:gamma};

\noindent
(2) assuming that \texttt{S} completely faded away before and after period S, we used the data for 4.9 years without period S and included only the 3FGL sources in the model (i.e.,\ without source \texttt{S}), which yielded the UL values denoted by superscript 'b' in Table \ref{tab:gamma}.

\noindent
The thus obtained ULs on $F_\mathrm{0.2-1\;GeV}$ differ by a factor of $\sim 4$, both values of the resulting UL on $\lambda_\mathrm{0.2-1\;GeV}$ are shown in Fig.\ \ref{fig:2}.
The UL on $F_\mathrm{1-10\;GeV}$ is much less model-dependent. In Fig.\ \ref{fig:1}b we use the UL obtained for the above case (1) with assumed $\Gamma=2.1$, other ULs for this energy range differ by at most 50\% and do not affect our conclusions.
We also note that adding source \texttt{S} in the model for the first three years has a negligible  effect, see cases 'cd' in Table \ref{tab:gamma}.
We also checked that our results are unchanged if we introduce additional sources to account for several residuals located at $\sim 1\degr$ to $\sim 4\degr$ from NGC 4151, seen in Fig.\  \ref{fig:tsmap}(a). 

We used the  recent dust-parallax distance measurement and the implied stellar-velocity-based mass (see Table \ref{tab:basic}),
both higher by $\sim 50$\% than assumed in previous works on high-energy emission from NGC 4151.
NGC 4151 shows a moderate X-ray variability with changes of the X-ray flux by up to a factor of $\sim 4$ between the dim and bright states. Using the adopted distance and $M$ values and the power-law fits  of \cite{2010MNRAS.408.1851L},  we found the intrinsic $\lambda_\mathrm{2-10\;keV} \simeq (0.5,1,2) \times 10^{-3}$ for the (dim, medium, and bright) state; the parameters of these three states are shown in Fig.\ \ref{fig:1}a). 
The average  {\it Swift}/BAT flux during the considered {\it Fermi}/LAT observations is higher by a factor 1.2 than the flux of the medium state, see Fig.\ \ref{fig:lc}b.

As we see in Figs \ref{fig:1}b and \ref{fig:2}, the ULs on $F_\mathrm{0.2-1\;GeV}$ and $F_\mathrm{1-10\;GeV}$
derived for NGC 4151 yield the tightest constraints on both $\lambda_\mathrm{0.2-1\;GeV}$ and $\lambda_\mathrm{1-10\;GeV}$ 
of the bright Seyfert 1 galaxies (although  IC 4329a has the lowest $L_\gamma / L_{\rm X}$ ratio in our sample).
For $\delta=10^{-3}$, model N overpredicts $\lambda_\mathrm{1-10\;GeV}$ by a factor of several, regardless of $a$ or $\beta$ values. 
Then, if heating of electrons is dominated by Coulomb interactions, the energy content of nonthermal protons is limited to $\eta_{\rm p} \la 0.2$.

For $\delta=0.5$, the model predictions, except for model N with $a=0$, are below the UL value.
However, as noted before, the much lower $L_{\gamma}$ predicted by large-$\delta$ models is accompanied by much lower $\tau$.
The X-ray spectrum of NGC 4151 is precisely measured, at least in the bright state, and allows testing this property.
Fits of the bright-state spectrum with the slab \texttt{compPS} model in \cite{2010MNRAS.408.1851L} give $\tau^{\rm PS} \simeq 1.3$. 
To compare it with our model predictions, we simulated the X-ray spectra for our models with $\lambda_\mathrm{2-10\;keV} \simeq 0.002$; we used the {\it INTEGRAL}/ISGRI response function and assumed the normalization corresponding to the bright state of NGC 4151. Then, the simulated spectra were fitted with the slab \texttt{compPS} model. 
For $\delta=10^{-3}$, the best fit has $\tau^{\rm PS} \simeq 1.1$
(model H$_{0.1}$ with $a=0.95$, $\beta=1$, $\dot m = 0.65$). For $\delta = 0.5$, all simulated spectra were fitted with $\tau^{\rm PS} \la 0.5$.
We conclude that although large-$\delta$ models are essentially unconstrained by the {\it Fermi} UL, they also appear to be inconsistent with the X-ray data. 

$\lambda_\mathrm{0.2-1\;GeV}$ predicted by model T, compared with with  the {\it Fermi} UL for the above case (1), disfavors models with both a large $a$ and a large $\beta$, in a manner consistent with NGC 7213. For case (2), however, the UL is above the model prediction. Stronger constraints can be obtained if NGC 4151 enters the flux level of the dim state for a longer period,
as at $\lambda_\mathrm{2-10\;keV} \simeq 10^{-3}$, $r_{\rm ph}$ is close to the extent of the $\gamma$-ray emitting region for thermal  protons.

\subsection{NGC 6814}

The $\gamma$-ray signal from the direction of NGC 6814 with TS $= 25.6$ was found by \cite{2012ApJ...747..104A}.
However, the source is not reported in 3FGL. It may not have passed the TS $>25$ criterion, assumed in 3FGL, as the above value is close to the limit.
Our analysis, using data for 6.4 years, confirms the signal with TS $\simeq 32.3,$
and our best-fit parameters for a power-law spectrum (see Table \ref{tab:gamma}) are 
approximately consistent  with those of \cite{2012ApJ...747..104A}.

In Fig.\ \ref{fig:1}a we show the parameters of the intrinsic X-ray emission from spectral fits to {\it Suzaku} observations in 
2011 \citep{2013ApJ...777L..23W}. 
Interestingly, the source is located in the region for an inefficient (thermal synchrotron) source of seed photons.
During this observation the source was 
in a relatively dim state, with the  average  {\it Swift}/BAT flux lower by a factor of $\sim 1.5$ than the average during the {\it Fermi}/LAT observations, see Fig.\ \ref{fig:lc}d.

As noted above, we considered two values of $M$ for NGC 6814. For both, the measured $F_\mathrm{1-10\;GeV}$ gives $\lambda_\mathrm{1-10\;GeV}$ exceeding the maximum value predicted for the hot flow (in models N with $\delta=10^{-3}$), which rules out an origin of the observed $\gamma$-ray signal in an inner hot flow. 
For the higher value of $M$, NGC 6814 has a similar $\lambda_\mathrm{1-10\;GeV}$ to the $\gamma$-ray-loud Seyfert 2 galaxies, a possible explanation is discussed in Sect. \ref{sect:discuss}.

\subsection{NGC 5548 and IC 4329a}

The likelihood analysis of the {\it Fermi}/LAT data does not reveal any significant signal around these two X-ray-bright Seyfert galaxies (in both TS $< 1$) and therefore we computed the ULs given in Table  \ref{tab:gamma}.

For NGC 5548, we show in Fig.\ \ref{fig:1}a the parameters of the power-law fits from Brenneman et al.\ (2012) for seven {\it Suzaku} observations in 2007. The average {\it Swift}/BAT flux during the 6.4 years of 
{\it Fermi}/LAT observations is equal to the average flux during the {\it Suzaku} observations, 
and we used the intrinsic $\lambda_\mathrm{2-10\;keV}$
from the fit of Brenneman et al.\ (2012) for the average {\it Suzaku} spectrum.  

For IC 4329a, we show in Fig.\ \ref{fig:1}a the parameters of the power-law fits from Brenneman et al.\ (2014) 
for lower and higher flux states during {\it Suzaku} and {\it NuSTAR} observations in 
2012. We used $\lambda_\mathrm{2-10\;keV}$ larger by a factor 1.3 to account for the difference between the average {\it Swift}/BAT flux for 6.4 years and the flux from measurement simultaneous with {\it Suzaku}/{\it NuSTAR} observations, see Fig.\ \ref{fig:lc}c.

Both sources are observed above $\lambda_\mathrm{ADAF, max}$ but within the range of approximate applicability of our model. During one observation, NGC 5548 is seen in the area of 
ADAF solutions with inefficient cooling. In the remaining observations both sources show interesting hints of modest hardening below
or softening above $\lambda_\mathrm{2-10\;keV} \simeq 0.002$ with increasing $\lambda_\mathrm{2-10\;keV}$.

For both objects the ULs on the $\gamma$-ray photon flux is lower by a factor of 2--3 times than for NGC 4151.
In addition, both sources have a similar $L_\gamma / L_{\rm X}$ ratio to NGC 4151 (for IC 4329a it is even lower). However, we see in   Fig.\ \ref{fig:1}b and Fig.\ \ref{fig:2} that they insignificantly constrain the hot flow models.

\subsection{NGC 7213}
\label{sect:7213}

Despite the large uncertainty in $M$,  NGC 7213 gives interesting  constraints on the hot flow model; crucially, it is observed at $\lambda_\mathrm{2-10\;keV} \simeq 10^{-4}$,  between the X-ray bright Seyfert 1 galaxies discussed above and  NGC 4258.
The likelihood analysis of the {\it Fermi}/LAT data does not reveal any significant signal around NGC 7213 (TS $< 1$),
and we computed the ULs given in Table  \ref{tab:gamma}.
In Fig.\ \ref{fig:1}a we show the parameters from a {\it Suzaku} observation in 
2006 \citep{2010MNRAS.408..551L} during which the average flux level was similar to the average during 6.4 years, see Fig.\ \ref{fig:lc}a. 

Figure\ \ref{fig:1}b shows that the derived UL for $F_\mathrm{1-10\;GeV}$ constrains the nonthermal energy content to 
$\eta_{\rm p} \la 0.1$ for $\delta=10^{-3}$. Models with $\delta=0.5$ predicts $\lambda_\mathrm{1-10\;GeV}$ below the {\it Fermi} limit, except for $a=0$ and $\eta_{\rm p}=1$.
However, the X-ray spectra predicted for $\delta=0.5$ appear to be qualitatively inconsistent with the spectrum measured in NGC 7213.
The spectra at $\lambda_\mathrm{2-10\;keV} \simeq 10^{-4}$ 
(the dotted line for $a=0.95$ and the overlying dashed line for $a=0$ in Fig.\ \ref{fig:0}c)
deviate significantly from a power-law shape,
whereas the {\it Suzaku} spectrum between  0.6 and 50 keV 
is described by a simple power-law  \citep{2010MNRAS.408..551L}.
The UL for $F_\mathrm{0.2-1\;GeV}$ rules out a high $\beta$ flow around a high-$a$ black hole.

\subsection{NGC 4258} 
The precisely determined parameters of this AGN made it an essential object for the development of the hot flow models \citep[e.g.,][]{1996ApJ...462..142L}. However, it has not been included in previous analyses of the {\it Fermi}/LAT data. At the position of NGC 4258 we found a weak $\gamma$-ray excess above the background, with TS = 9.4.
A likelihood analysis of the signal with \texttt{gtlike} gives $\Gamma = 2.5 \pm 0.2$, $F \simeq (2.8 \pm 1.2) \times 10^{-9}$ ph s$^{-1}$ cm$^{-2}$.
Despite the weak residual, the derived ULs on the photon flux yield the tightest constraint on $\lambda_\mathrm{1-10\;GeV}$ and $\lambda_\mathrm{0.2-1\;GeV}$ of all considered objects, see Table \ref{tab:gamma}.

In Fig.\ \ref{fig:1}a we show parameters of the analysis reported
by \cite{2009ApJ...691.1159R} of the {\it Suzaku} observation in 
2006, who also noted that during this observation NGC 4258 increased its intrinsic X-ray luminosity by a factor of $\sim 2$ relative to the average from {\it Swift}/BAT measurements. We used $\lambda_\mathrm{2-10\;keV}$
from the {\it Suzaku} fit  reduced by a factor of 2.

At  $\lambda_\mathrm{2-10\;keV} \simeq 10^{-5}$, the 0.2--1 GeV emission is weakly affected by $\gamma \gamma$ absorption,
and we can observe radiation from the ergospheric region (see Sect. 2). 
However, as Fig.\ \ref{fig:2} shows, the strongest expected signal is only at the level of the {\it Fermi} UL.
 
For $\delta=10^{-3}$, model N predicts a flux strongly exceeding the derived UL for $F_\mathrm{1-10\;GeV}$, ruling out $\eta_{\rm p} \sim 1$. For $\delta = 0.5$, the predicted $\gamma$-ray flux is an order of magnitude below the UL value.
However, the corresponding X-ray spectrum (bottom dashed line in Fig.\ \ref{fig:0}c) seems to rule out  this (large-$\delta$) case if the X-ray radiation observed from NGC 4258
is produced by thermal Comptonization.

\subsection{Circinus, NGC 1068, and NGC 4945}
\label{sect:sy2}

For the Seyfert 2 galaxies we used the parameters reported in 3FGL, which are approximately consistent ($F_{\rm 1-10\;GeV}$ within 30\%) 
with previous analyses \citep{2010A&A...524A..72L, 2012ApJ...755..164A,2013ApJ...779..131H}.

Of these three AGNs, only NGC 4945 shows rapid variability in the hard X-rays, which indicates that it 
is a transmission-dominated source. We can therefore directly probe the nuclear emission. In Fig.\ \ref{fig:1}a we show fits from \citet{2014ApJ...793...26P} for {\it NuSTAR} observations in 2013. The average  {\it Swift}/BAT flux during the four years of {\it Fermi} observations used in 3FGL agrees with the average flux during the {\it NuSTAR} observations.

In NGC 1068, the direct X-ray emission of is completely obscured along our line of sight and we only see the reflected component. Then, the assessed intrinsic emission is strongly model dependent. We used  $\lambda_\mathrm{2-10\;keV}$
corresponding to the best-fit of \cite{2014arXiv1411.0670B}. This assessment, based on the observed reflected component, gives the level of emission from the active nucleus averaged over a long time, possibly over hundreds of years, because a significant fraction of reflection arises at a $\sim 100$ pc scale \citep[see][]{2014arXiv1411.0670B}. 
Similarly, the X-ray emission from Circinus is reflection dominated; we used results from \cite{2014ApJ...791...81A}. NGC 1068 and Circinus are not shown in Fig.\ \ref{fig:1}a because the estimates for their intrinsic $\Gamma_{\rm X}$
are model dependent.

Figure\ \ref{fig:1}b shows that all three $\gamma$-ray-loud Seyfert 2 galaxies agree remarkably well  after scaling by their central black hole masses, with $\lambda_\mathrm{2-10\;keV} \simeq 0.02$ and $\lambda_\mathrm{1-10\;GeV} \simeq 3 \times 10^{-5}$ in all three.
We also note that their $L_{\rm 1-10\;GeV}/L_{\rm 2-10\;keV}$ ratio does not exceed the limit on the luminosity ratio  in other AGNs, except for that corresponding to the UL in IC 4329a for $\Gamma=2.7$.

\subsection{Centaurus A}

The interpretation of the high-energy spectrum of Cen A has some open questions.
Its X-ray emission may contain contribution from both the jet and the accretion flow. 
The latter origin is supported by the thermal-like cutoff, claimed in \cite{2011A&A...531A..70B}, and by 
the location in the area of the $\lambda_\mathrm{2-10\;keV}$--$\Gamma_{\rm X}$ plane occupied by Seyfert galaxies. 
In Fig.\ \ref{fig:1}a we show the parameters from three {\it Suzaku} observations in 2009 \citep{2011ApJ...743..124F}.
\cite{2011ApJ...743..124F} also noted hints for a harder power-law component, with $\Gamma_{\rm X} < 1.6$, weakly contributing below 100 keV, which may represent emission from the jet.
The main contribution to the $\gamma$-ray emission of Cen A most likely comes from the jet. However, \cite{2013ApJ...770L...6S} found evidence for a second component, an order of magnitude less luminous than the main $\gamma$-ray component. This weaker component has a hard spectrum and dominates above $\sim 4$ GeV. Assuming that this component extends down to 1 GeV with the same slope, we find $\lambda_\mathrm{1-10\;GeV} \simeq 1.7 \times 10^{-6}$. Based on a comparison with predictions of model 
 H$_{0.1}$ , this component can be emitted from a flow with $\delta=10^{-3}$ and $\eta_{\rm p} \simeq 0.1-0.2$ (slightly dependent on plasma magnetization). 

The {\it Swift}/BAT flux level during the 2009 {\it Suzaku} observations
is equal to the average flux during the {\it Fermi}/LAT data-taking period used for the analysis in \cite{2013ApJ...770L...6S}.

\section{Nonthermal electrons}

\label{nthel}

Here we consider effects related with nonthermal electrons that
are produced by 
various mechanisms, specifically, their contribution to the observed $\gamma$-ray luminosity and to the optical depth.

\subsection{Pion decay}

If nonthermal protons are present, electrons with $\gamma \ga 10^3$ may be produced by charged pion decay. The nonthermal Compton radiation of these electrons may extend to the LAT energy range.
The power injected in $\pi^{\pm}$ decay electrons is similar to the luminosity in $\pi^0$ decay $\gamma$-rays.
However, electrons with $\gamma \ga 10^3$ lose most of their energy in synchrotron radiation, therefore their nonthermal Compton emission in the $\gamma$-ray range is always much weaker than the $\pi^0$ decay component.

The density of $\pi^{\pm}$-decay $e^{\pm}$, determined by the pair equilibrium condition \citep[cf.][]{2015ApJ...799..217N}, is at least two orders of magnitude lower than the density of the ionization electrons.
The production rate of $\pi^0$ is similar to that of $\pi^{\pm}$, therefore $e^{\pm}$ pairs produced by absorption of $\pi^0$-decay $\gamma$-rays give at most a similar contribution to the total density as those from  $\pi^{\pm}$-decay. Production of further generations of $e^{\pm}$ pairs is unlikely for the luminosities described by our model, see below.  
Overall, the products of hadronic processes 
contribute negligibly to the optical depth. 

If a strong $\gamma \gamma$ absorption affects the $\pi^0$-decay $\gamma$-rays, the synchrotron radiation of thus produced $e^{\pm}$ pairs increases the seed photon input, again at most by a factor of $\sim 2$. We neglected 
this effect in computations presented in this work; we checked that this twice higher seed photon input yields an electron temperature that is reduced by up to 10\% and only insignificantly affects our results. 

\subsection{Direct acceleration}

\label{sect:accel}

We briefly discuss constraints on the acceleration of electrons resulting from 
the {\it Fermi}/LAT ULs. For the results presented in this section we do not use the simplified description (i.e., with a monoenergetic injection) of the accelerated electrons, and we assume that they are accelerated to the power-law distribution with the acceleration index $s_{\rm e}$.
As above, the amount of nonthermal electrons is parametrized by $\eta_{\rm e}$. Specifically, the power used for the nonthermal acceleration is $Q_{\rm acc} = \eta_{\rm e} \delta Q_{\rm diss}$, where $Q_{\rm diss}$ is the total power dissipated in the flow. 

Conditions in the inner parts of hot flows surrounding supermassive black holes allow for acceleration of electrons to GeV energies; for instance,\ using Eq.\ (7) from \cite{2009MNRAS.394L..41Z}, we found the maximum Lorentz factors to be limited by the synchrotron losses of $\gamma_{\rm max} \sim 10^5 - 10^6$ for the typical magnetic field strength of $B \sim (10^2 - 10^4)$ G.
The synchrotron emission of these electrons extends up to $\sim 10$ MeV, so they can produce photons in the LAT energy range only by Compton scattering. 
We implemented the description of nonthermal Compton emission in our MC code and performed simulations with $\gamma_{\rm max} = 2 \times 10^4$, for which the nonthermal Compton spectrum extends up to $\sim 10$ GeV. 

The $\gamma$-ray luminosity produced by  the nonthermal Compton scattering primarily depends on $\eta_{\rm e}$ and
 $s_{\rm e}$.  We found that for $s_{\rm e} \ga 2.4$ and $\eta_{\rm e}=1$,
the values of  $\lambda_{\rm 0.2-1\;GeV}$ and $\lambda_{\rm 1-10\;GeV}$ for the nonthermal Compton are below any UL derived from the LAT data for objects  
with $\lambda_{\rm 2-10\;keV} \la 10^{-3}$. 
For $s_{\rm e} \la 2.4$, the $\gamma$-ray Compton component for $\eta_{\rm e}=1$ exceeds the ULs for NGC 4258, NGC 7213 and NGC 4151, then, for such $s_{\rm e}$
the fraction of accretion power used for nonthermal acceleration is constrained. We found for instance that for $s_{\rm e} \simeq 2$ it is limited to $\la 0.05 Q_{\rm diss}$ (corresponding to $\delta=0.5$ and   $\eta_{\rm e} \la 0.1$).

In our computations we assumed that synchrotron radiation emitted below the self-absorption frequency is thermalized, as described in several works on the synchrotron boiler mechanism \citep[e.g.,][]{2009MNRAS.392..570M}, meaning that\ the effective heating of thermal electrons
is higher than $(1-\eta_{\rm e}) \delta Q_{\rm diss}$.
The power in the self-absorbed radiation is $\ga 0.2 Q_{\rm acc}$ for $s_{\rm e} \ga 2$. For $s_{\rm e} = 2.6$, it is $\sim 0.8 Q_{\rm acc}$.
Then, if $\delta$ is large, the heating of thermal electrons is always strong, with the synchrotron boiler dominating for $\eta_{\rm e} \sim 1$.
This also implies that our conclusions regarding the allowed values of $\eta_{\rm e}$ are weakly dependent on other parameters of the flow. Although  $Q_{\rm diss}$ shows a significant dependence\ on $a$, for example, 
both the nonthermal Compton $L_{\gamma}$ and the thermal Compton $L_{\rm X}$ scale roughly linearly with $Q_{\rm diss}$ (the latter either through the synchrotron boiler or direct MHD heating), making the $L_{\gamma}/L_{\rm X}$ ratio weakly dependent on the other parameters.

\subsection{$\gamma  \gamma$  absorption}

Secondary $e^{\pm}$ pairs, produced by $\gamma  \gamma$  absorption, in general affect radiative properties of flows in the considered range of   $\lambda_{\rm 2-10\;keV}$ only insignificantly. The first generation of pairs contributes at most a few per cent to the Thomson optical depth. Copious creation of secondary pairs through electromagnetic cascades requires (1) the Compton process to dominate over synchrotron, and (2) Compton photons produced above the threshold for pair production. Typically, at least one of these conditions is not fulfilled. In the innermost parts of the flow,
Compton energy losses are comparable to synchrotron losses  only for electrons with Lorentz factors not exceeding a few hundred.
At higher energies, synchrotron losses strongly dominate as a
result of the Klein-Nishina decline in the Compton scattering rate, hence, the development of a cascade is suppressed.
In turn, the lower energy electrons, for which the more efficient Compton losses are mostly due to interactions with X-ray photons from thermal Comptonization, mainly produce photons with  $E \la 10$ MeV. However, at $\lambda_{\rm 2-10\;keV} \la 10^{-3}$, the flows are mostly transparent to such photons. 
Therefore, we do not expect a significant contribution of secondary pairs to the Thomson depth of the flow. 
We note, however, that at $\lambda_{\rm 2-10\;keV} \ga 10^{-3}$ the innermost region should be opaque to photons with $E \ga 1$ MeV, and in this regime the secondary pairs may then contribute considerably to the optical depth.

We also note that spatially extended pair cascades, such as discussed\ by \cite{2009MNRAS.394L..41Z}, for instance, are unlikely to strengthen the observed $\gamma$-ray flux. Such extended cascades can only develop in the region with $\tau_{\gamma \gamma} \sim 1$. In a hot flow,  $\tau_{\gamma \gamma}$ increases quickly toward the center, see Fig. 7 in N13, therefore the free path of $\gamma$-rays produced within $r_{\rm ph}$ is typically very short. Furthermore, as noted above, the absorbed energy is lost to synchrotron radiation rather than used for cascade development.

\section{Discussion}

\label{sect:discuss}

We first briefly comment on the X-ray spectral properties of
the considered AGNs with $\lambda_{2-10} \la 10^{-3}$.
At these luminosities, the seed photons for Comptonization must be produced internally to the flow,
and nonthermal synchrotron radiation is an almost obvious process providing the amount of seed
photons needed to explain the AGN data \citep{2014SSRv..183...61P,2015ApJ...799..217N}. If $\delta$ is small, the nonthermal electrons are provided by the decay of charged pions, unless both $a$ and $\beta$ are small and protons are thermal, see Sect. \ref{sect:t}.   We emphasize that the weakness of $\gamma$-ray signals related with $\pi^0$ decay does not contradict  a copious injection of nonthermal electrons by $\pi^\pm$ decay because the most efficient pion production occurs in the region opaque to $\gamma$-rays.
If $\delta$ is large, the $\pi^\pm$ decay does not provide a sufficient amount of nonthermal electrons because the flow density is too low (here inefficient $e^\pm$ production is strictly related with very low level of $\gamma$-ray activity).
Then,  direct acceleration must take place if $\delta$ is large. 
These models (i.e.,\ with either a significant $\pi^\pm$ production or  $\eta_{\rm e} \ga 0.1$)  fill the upper gray region in Fig.\ \ref{fig:1}a.
Typical Seyfert spectra clearly agree with these solutions.

Solutions with  seed photons from thermal synchrotron alone (including models with $\delta=0.5$ and $\eta_{\rm e}=0$, as well as  model T with $\delta=10^{-3}$, $a=0$ and $\beta=1$) fill the lower gray region. Obviously, intermediate cases are also possible, for instance,\ with $\delta=0.5$ and $\eta_{\rm e} \la 0.01$. 
Seyferts are occasionally observed with parameters corresponding to inefficient cooling 
(NGC 6814 and one of NGC 5548 data points). These harder-than-typical spectra may result from a decline of nonthermal electron injection. For large $\delta$ it may correspond to the decrease of $\eta_{\rm e}$, and for small $\delta$
to the decrease of $\beta$ (again, for small $a$ and thermal protons).

Observational assessment of nonthermal acceleration processes in astrophysical plasmas is an essential question. 
Hot flows in which ultra-relativistic acceleration of protons takes place can be considered a source of neutrinos  detected by {\it IceCube} \citep[e.g.,][]{2015ApJ...806..159K}. 
Accretion flows should also be strong acceleration sites according to hybrid thermal and nonthermal Comptonization models 
\citep[e.g.,][]{2009ApJ...698..293V,2009MNRAS.392..570M,2011MNRAS.414.3330V} that are widely applied to explain X-ray spectra of accreting black holes; here at least $\sim 50$\% of the energy must be provided to electrons by nonthermal processes if  thermalization of electrons via Coulomb collision and synchrotron self-absorption is taken into account \citep{2013MNRAS.430..209D}.
Interestingly, these models are typically fitted with $s_{\rm e} \ga 2.4$.
As discussed in Sect. \ref{sect:accel}, the $\gamma$-ray luminosities from nonthermal Compton emission of flows with such $s_{\rm e}$ are below the LAT ULs derived in Sect. \ref{sect:obs}   for Seyfert galaxies, even if
$\eta_{\rm e}=1$.

Observational evidence of the presence of non-thermal particles in accretion flows includes
detection of MeV tails \citep[e.g.,][]{2010ApJ...717.1022D,2012MNRAS.423..663Z}
and patterns of optical/infrared evolution \citep{2014MNRAS.445.3987P}, both observed in stellar black-hole systems. 
However, these features are produced by nonthermal electrons, which may come from pion decay, and in principle, they do not require direct acceleration. 

Observations in the GeV range provide the means to directly measure the efficiency of proton acceleration in hot flows.
The obtained ULs for $F_\mathrm{1-10\;GeV}$ in NGC 4258, NGC 7213 and NGC 4151 constrain the fraction of accretion power used for relativistic acceleration of protons to at most $\eta_{\rm p} \simeq 0.1$ (if $\delta \la 0.1$). 
Remarkably, these three AGNs allow probing various parts of the flow because at $\lambda_\mathrm{2-10\;keV} \sim 10^{-5}$, $10^{-4}$  and $10^{-3}$, radiation produced at a few, ten, and several tens of $R_{\rm g}$, respectively, dominates the observed 1--10 GeV flux.
The component dominating the core emission of Centaurus A above  $\sim 4$ GeV may come from a hot flow in which $\eta_{\rm p} \simeq 0.1 - 0.2$. In this scenario, its $\eta_{\rm p}$  should be higher at least by a factor of $\sim 2$  than in NGC 7213 (see Fig.\ \ref{fig:1}b), which has similar $\lambda_\mathrm{2-10\;keV}$. This might reflect differences between flows powering radio-loud  and radio-quiet  AGNs. We also note that in this hot flow scenario for Centaurus A, the observed $\gamma$-rays (above 4 GeV) provide information about the region close to the jet formation site\footnote{according to the radio imaging of M 87, e.g. \cite{2012Sci...338..355D}, and assuming that the sizes of such sites are common in AGNs}.

We now briefly discuss hard spectral states of black hole binaries in which
hot accretion flows, similar to those powering AGNs, are most likely present.
In the innermost parts of flows powering both classes of objects the same physical processes
are expected. This view is supported by the similarity between the X-ray spectra of Seyferts 
and black hole binaries in the hard spectral states \citep[e.g.,][]{1996A&AS..120C.553Z},
with slight differences simply explained by scaling of seed photon energies with the black hole mass 
\citep[cf.][]{2015ApJ...799..217N}. These differences do not affect the $\gamma$-ray emission, 
and the relation between $\lambda_{\rm 2-10~keV}$ and $\lambda_{\rm 0.2-1~GeV}$ or $\lambda_{\rm 1-10~GeV}$
shown in Figs \ref{fig:1}--\ref{fig:2} is probably also relevant for flows around stellar-mass black holes.

An estimate of the $\gamma$-ray activity level, which can be compared with the hot flow model, has only been obtained for the hard state of Cyg X-1 by \cite{2013MNRAS.434.2380M}.
 Figure\ \ref{fig:1}b shows that it gives a slightly tighter constraint than Seyfert galaxies, 
limiting $\eta_{\rm p}$ to at most 
a few per cent for $\delta=10^{-3}$, and at most $\sim 10$\% for $\delta=0.5$ and $a=0$. 
A significant nonthermal proton component, with $\eta_{\rm p} \sim 0.5$, is only allowed if $\delta \sim 0.5$ and $a$ is high.
This finding places interesting constraints on hybrid models (see above) if protons are accelerated with at least a similar efficiency to electrons (as expected in many acceleration scenarios).
It may be also interesting to note that similar values of $\lambda_{\rm 1-10~GeV}$
correspond to the Cyg X-1 hard state ($\simeq 6 \times 10^{-7}$) and our \texttt{gtlike} fits to signals at the positions of NGC 4151
($\simeq 5 \times 10^{-7}$) and NGC 4258 ($\simeq 3 \times 10^{-7}$), although the significance of the $\gamma$-ray signal is rather weak (TS = 15.6) in Cyg X-1 and marginal in NGC 4258 and NGC 4151.

Finally, we comment on the $\gamma$-ray emission of NGC 4945, NGC 1068, and Circinus. All three exhibit both an  AGN and starburst 
activities. NGC 4945 and NGC 1068 approximately follow some multiwavelength correlations established in other star-forming galaxies  \citep{2012ApJ...755..164A}, which suggests that their $\gamma$-ray emission may be related to the latter activity.
However, Circinus does not agree with these correlations, furthermore, its $L_{\gamma}$ exceeds the calorimetric limit (where all cosmic rays produced by supernovae interact with interstellar matter), see \cite{2013ApJ...779..131H}.
Furthermore, \cite{2010A&A...524A..72L} noted that $L_{\gamma}$ measured in NGC 1068 exceeds by a factor of $\sim 10$ the level expected for the total gas mass and supernova rate estimated in this galaxy, which led them to suggest that the $\gamma$-rays  may be produced in a jet. 
Similarly, \cite{2014ApJ...780..137Y} found that their detailed starburst model underestimates the observed $\gamma$-ray flux and overestimates the radio flux for NGC 1068, and they argued that these problems would be resolved if the AGN were the primary source of $\gamma$-rays.
This shows that the origin of $\gamma$-rays still seems rather unclear.

Seyfert galaxies show a strong correlation between nuclear star-formation and the AGN luminosity \citep[e.g.,][]{2012ApJ...746..168D}.
Masses of supermassive black holes in Circinus, NGC 1068, and NGC 4945  are precisely determined by H$_2$0 megamaser measurements.
Their values imply an intriguing similarity of the values of both $\lambda_\mathrm{2-10\;keV}$ and $\lambda_\mathrm{1-10\;GeV}$ in all three objects.
The question then arises whether their $\gamma$-ray emission might be related with accretion processes in their active nuclei. For the intrinsic $\lambda_\mathrm{2-10\;keV} \simeq 0.02$, the
geometry of accretion flow is very uncertain. 
It cannot be studied through Fe K$\alpha$ line distortions because the X-ray spectra of all three AGNs are completely absorbed 
below 10 keV.
Some analogies with black hole binaries, in which (1) luminous hard states have been observed at such $\lambda_\mathrm{2-10\;keV}$ \citep[e.g.,][]{2004MNRAS.351..791Z},
(2) softening of intrinsic X-ray slopes above $\lambda_\mathrm{2-10\;keV} \simeq 10^{-3}$ is observed that
closely resembles the behavior seen in  NGC 4945 (see, e.g.,\ Fig. 1 in \cite{2015MNRAS.447.1692Y}),
and (3) strong thermal components are seen at $\lambda_\mathrm{2-10\;keV} \sim 0.01,$ indicating that the cold disk extends down to $r_{\rm tr} \la 10$ 
\citep[e.g.,][]{2009MNRAS.396.1415C}
suggest that inner hot flows may be present in these AGNs, but only within the innermost several $R_{\rm g}$.
This value of $r_{\rm tr} \sim 10$ is tentatively shown in Fig.\ \ref{fig:1}a.
For such a small, strongly energized  (the main release of gravitational energy occurs in this region) and relatively dense flow, we cannot exclude the possibility that a sufficient $\gamma$-ray flux is produced in an  outer layer and   escapes avoiding $\gamma \gamma$ absorption (taking into account a strong local anisotropy of the X-ray radiation field, see Sect. 2).
Any stronger conclusions for this scenario would be premature; a self-consistent model, compatible with the X-ray data, needs
to be developed before a quantitative assessment of the $\gamma$-ray flux can be made.

We do not expect direct $\gamma$-ray signals from central parts of accretion flows in black hole binaries even if they
have the same geometry at this  $\lambda_\mathrm{2-10\;keV}$.
Any emission in the GeV range would be absorbed in interaction with soft X-ray emission of a cold disk 
extending to small $r$, see \cite{1993A&A...278..307B} for quantitative estimates.
In AGNs, a UV-emitting cold disk only attenuates $\gamma$-rays at $\sim 100$ GeV.

Another possibility involves escape of protons accelerated in the flow and their interaction with circumnuclear matter.
About 0.1\% of the accretion power would need to be carried away
by escaping protons to explain the $\gamma$-ray measurements. 
This would also allow explaining the $\gamma$-ray signal in NGC 6814 by the same emission mechanism,
assuming that (1) its supermassive black hole has $M \simeq 2 \times 10^{7}M_{\odot}$ (the higher of the two values considered above), for which 
its $\lambda_\mathrm{1-10\;GeV}$ is similar to those of NGC 4549, Circinus, and NGC 1068; and (2) protons were injected into the circumnuclear 
region during an earlier phase of higher (by almost 2 orders of magnitude relative to current) activity of NGC 6814,
in the scenario similar to that proposed for the Galactic center \citep[e.g.,][]{2011ApJ...726...60C}.
In this scenario, an energy-dependent diffusive loss of protons may explain the slightly softer $\gamma$-ray spectrum of NGC 6814.

\section{Summary}

Hot flows are supported by the proton pressure, and 
the production of pions in collisions of the energetic protons should be their generic property.
If this is true, a lack of detection of a related $\gamma$-ray signal could be regarded as an argument against this class of models.
However, the expected flux depends on several parameters, and in the extreme case of a strongly magnetized flow 
around a nonrotating black hole, the conditions for pion production are not achieved if protons are thermal.

We compared the prediction of the model with several well-studied AGNs, for which the available data allowed a robust determination 
of the nuclear luminosity scaled by the Eddington value. At the positions of NGC 4151 and NGC 4258, we found
weak residuals, whose nature (background fluctuation or an actual emission) could not be assessed because of their
low statistical significance.

The highest $\gamma$-ray luminosities are predicted if electrons are heated  mostly by Coulomb interactions.
If additionally most of the accretion power is used for the relativistic acceleration of a small fraction of protons, the highest
level of $\lambda_\mathrm{1-10\;GeV} \simeq 10^{-5}$ is expected for $\lambda_\mathrm{2-10\;keV}$ between  $\sim 10^{-5}$ and $10^{-3}$.
This exceeds  the {\it Fermi} limit on $\lambda_\mathrm{1-10\;GeV}$ for NGC 4258, NGC 7213, and NGC 4151 at least by a factor of several.
Thus, the {\it Fermi} ULs provide an interesting constraint on
the MHD processes that convert the accretion power into the kinetic
energy of protons: A rather uniform heating of all protons would
need to take place.
Given this constraint,
 we used $\lambda_\mathrm{0.2-1\;GeV}$ predicted for thermal distribution 
of protons to rule out a flow with a high (equipartition) value of $\beta$ around a high-$a$ black hole for NGC 7213 and 
and NGC 4151; for the latter this result is model dependent.

If a large part of the accretion power is used for a direct heating
or acceleration of electrons, 
the expected $\gamma$-ray fluxes are below the {\it Fermi} ULs, except for NGC 4151 and NGC 7213, for which
 $\eta_{\rm p} \ga 0.5$ was ruled out with a small $a$ (yielding some constraints on the acceleration efficiency,
which may be relevant for hybrid models).
However, this (large $\delta$) class of models appears to be
disfavored by the predicted properties of X-ray spectra,
which significantly deviate from a power law at $\lambda_\mathrm{2-10\;keV} \la 10^{-4}$ 
and are described by (too) small $\tau$ at larger $\lambda_\mathrm{2-10\;keV}$.

The $\gamma$-ray-loud Seyfert 2 galaxies radiate at a much higher Eddington ratio than other nearby AGNs,
possibly as a result of enhanced fueling that is related with starburst activity in their nuclear regions.
At this  $\lambda_\mathrm{2-10\;keV}$, the transition between the hot flow and the cold disk possibly takes place within the central $\sim 10 R_{\rm g}$. Whether it creates conditions suitable for the escape of $\gamma$-rays or relativistic protons remains an open question.

\begin{acknowledgements}
We thank F.\ Longo for help with the {\it Fermi}/LAT data analysis, P.\ Lubi\'nski for help with the {\it INTEGRAL} software and the referee for motivating us to include the results for nonthermal Compton emission. 
We made use of data and software provided by the Fermi Science Support Center, managed by the HEASARC at the Goddard Space Flight Center, and Swift/BAT transient monitor results provided by the Swift/BAT team.
RW was supported by the Polish MNiSW grant for young scientists B1511500001045.02.
AN was supported by the Polish NCN grant DEC-2011/03/B/ST9/03459.
FGX was in part supported by the Strategic Priority Research Program "The Emergence of Cosmological Structures" of CAS (grant XDB09000000), and the Natural Science Foundation of China (grants 11203057, 11133005).
\end{acknowledgements}

\bibliographystyle{aa}
\bibliography{26621_ap2}
\end{document}